\documentclass[nofootinbib,aps,pre,twocolumn,showpacs,groupaddress,floatfix]{revtex4-1}

\usepackage{amsmath}
\usepackage{booktabs}
\usepackage{mathrsfs}
\usepackage{graphicx}
\usepackage{fullpage}
\usepackage[caption=false]{subfig}
\usepackage{mathtools}
\usepackage{hyperref}
\usepackage[table]{xcolor}
\usepackage{tikz}
\usetikzlibrary{external}
\tikzsetexternalprefix{tikz_cache/}

\DeclarePairedDelimiterX{\infdivx}[2]{(}{)}{%
  #1\;\delimsize\|\;#2%
}

\renewcommand{\l}[1]{\mathcal{#1}}

\ifdefined\DRAFT
\newcommand{\alert}[1]{\textbf{\textcolor{red}{#1}}}
\newcommand{\hcwu}[1]{\textcolor{blue}{hcwu: #1}}

\else
\newcommand{\alert}[1]{}
\newcommand{\hcwu}[1]{}
\fi

\begin{document}

\title{Correlated structural evolution within multiplex networks}

\author{Haochen Wu}
\email{hcwu@ucdavis.edu}
\affiliation{%
    Department of Computer Science \\
    University of California, Davis, CA 95616, USA
}

\author{Ryan G. James}
\email{rgjames@ucdavis.edu}
\affiliation{%
	Complexity Sciences Center\\
    Department of Physics \\
    University of California, Davis, CA 95616, USA
}

\author{James P. Crutchfield}
\email{chaos@ucdavis.edu}
\affiliation{%
	Complexity Sciences Center\\
    Department of Physics \\
    University of California, Davis, CA 95616, USA
}

\author{Raissa M. D'Souza}
\email{rmdsouza@ucdavis.edu}
\affiliation{Complexity Science Center}
\affiliation{Department of Computer Science}
\affiliation{%
    Department of Mechanical and Aerospace Engineering\\
    University of California, Davis, CA 95616, USA
}
\affiliation{%
    Santa Fe Institute, 1399 Hyde Park Road, Santa Fe, NM 87501, USA
}
\begin{abstract}
Many natural, engineered, and social systems can be represented using the framework of a layered network, where each layer captures a different type of interaction between the same set of nodes. The study of such \emph{multiplex networks} 
is a vibrant area of research.  
Yet, understanding how to quantify the correlations present between pairs of layers, and more so present in their co-evolution,
is lacking.  Such methods would enable us to address fundamental questions involving issues such as function, redundancy and potential disruptions. 
Here we show first how the edge-set of a multiplex network can be used to construct an estimator of a joint probability distribution describing edge existence over all layers.  We then adapt an information-theoretic measure of general correlation called the conditional mutual information, which uses the estimated joint probability distribution, to quantify the pair-wise correlations present between layers.  The pair-wise comparisons can also be temporal, 
allowing us to identify if knowledge of a certain layer can 
provide additional information 
about the evolution of another layer.
%
We analyze datasets from
three distinct domains---economic, political, and airline 
networks---to demonstrate how pair-wise correlation in structure and dynamical evolution between layers can be identified and show that anomalies can serve as potential indicators of major events such as shocks. 
\end{abstract}

\maketitle

\ifdefined\DRAFT
\section*{To do}

\alert{Give interpretation of CMI applied to example data sets. Now, mostly
the manuscript simply includes heat map figures, but does not interpret them
and draw conclusions about what is learned.}

\alert{Adopt dynlearn.sty for ease of journal submission.}

\fi

\section{Introduction}
\label{sec:intro}

Over the last two decades, network analysis has become a useful tool for 
understanding social, biological, physical, and engineered complex systems~\cite{Newman10}. At its most basic, a network is a set of nodes and edges, where edges denote pairwise interactions between nodes. 
Although this ignores many details and higher-order interactions, such as multivariate dependencies beyond dyadic, this network approximation has yielded important insights into the formation and dynamics of
complex systems. 
Beyond a simple network, many real systems are composed of layers of individual networks, ranging from multimodal transportation networks to the Internet protocol stack, 
and interpreting such a system as a single-layer network of homogeneous interactions is often an
oversimplification~\cite{Zanin15}. 
In many instances each network layer contains the same set of nodes, but the edges in each distinct layer represent a distinct type of interaction between nodes.  We use the terminology \emph{multiplex network} to describe such a system. In the transportation setting, a multiplex network can be constructed where the nodes are geographic locations and each layer represents connectivity of a different transport type between locations, such as automobile, airplane, train, passenger ship, etc.  In recent years, there has been a vibrant study of multiplex networks~\cite{Kivela2014,Boccaletti2014,bianconi2018multilayer}. 

Generally, layers within a multiplex network are not independent. Consider
the multiplex transportation network above. 
Due to geographic constraints, there will be little overlap between edges in the automobile and in the passenger ship layers, so
the presence of an edge in one layer implies the likely absence in the other. 
The straightforward  
approach of estimating the
edge overlap rate between two network layers (\textit{i.e.}, the frequency at which two layers contain an
edge between the same pair of nodes) and then comparing that to a random network
null model demonstrates that correlation between layers are commonly found in multiplex
networks~\cite{Battiston14}. Yet, basic edge overlap does not account for many important features such as anti-correlation
in the location of edges which may depend on the characteristics of the system. For instance, the layers can variously either
cooperate or compete with one another, likewise their structures can be complementary or redundant, which all influence edge overlap.
Beyond static measures, 
measures are also needed to quantify correlations present during evolution. 



Such measures would allow a more
nuanced understanding of the dynamics underlying multiplex networks. For instance, do some layers evolve independent of all others? 
Can we find temporal correlations indicating that one layer influences the evolution of another? 
In addition, we can use these measures to understand real-world multiplex networks across domains.  For instance, there are principled arguments based on political and economic considerations 
that alliance treaties between nation states are related to their trade relationships~\cite{mansfield1997alliances}.
Quantitative measures would allow us to establish this explicitly and also identify if 
specific types of goods are more dependent on the alliance than others.
This, in turn, may reveal potential trade interventions that can impact the stability of alliances.
Likewise, while diplomatic disputes between nations may lead to war, which among the various classes of disputes is most influential? 
In a different realm, airline companies compete and cooperate with each other, but to what extent 
does one company's decision influence another's? 

Here we develop measures to quantify the correlation present between a pair of layers in a multiplex network as well as to quantify the correlation present in the co-evolution between a pair of layers.  Our primary contributions are two-fold. First, we develop a method for
constructing an estimator of the joint probability distribution describing the simultaneous existence of edges across layers of the multiplex network and also the discrete-time evolution of the edges. 
Second, using the joint probability distributions, we develop a conditional mutual information measure
that quantifies the extent to which one layer influences another, 
elucidating the correlated structural evolution of layers. 
We apply this method to a variety of empirical datasets from 
airline, political, and trade networks.
Our empirical studies reveal 
nontrivial 
relationships between layers
with some pairs of layers evolving in a more correlated manner
than other pairs. The method introduced here allows us to explore these
interlayer relationships and also to determine the temporal order of changes in
different layers. 

The development is organized as follows. Section~\ref{sec:related_work} discusses previous
related work. Section~\ref{sec:method} describes how we characterize a multiplex
network and its correlated internal structural evolution with a joint probability
distribution, and it also defines our conditional mutual information measure and 
tests for statistical significance. Section~\ref{sec:application} then applies
our method to multiple datasets, demonstrating its utility. Finally,
Sec.~\ref{sec:conclusion} concludes by summarizing potential limitations and
promising future directions.

\section{Related work}
\label{sec:related_work}

Techniques from information theory offer quantitative methods to extract correlations in time series data, but have not yet been extended to the multivariate setting required to analyze large networks.
That said, techniques from network science for analyzing multiplex networks provide insight into similarity of dynamics on layers or provide measures of ensemble properties, yet they do not extract correlations in structure and structural evolution. 
Here, we review these works organized along the two broad categories of approaches.  

\subsection{Information theory}

A multitude of existing techniques can be applied to time series data 
to quantify the correlations that are present,
including delay-coordinate embedding, Granger
analysis, and time-delayed mutual information~\cite{kantz2004nonlinear}.
Unfortunately, these techniques have yet to be extended to the multivariate
setting required for networks beyond a few nodes in size. 
Extending these informational measures to multivariate cases is an active area of current 
research~\cite{james2016information,james2017multivariate} which remains an open question. For
example, transfer entropy~\cite{Schreiber00,Barnett2009} measures the
time-asymmetric information shared between two random processes. And, it
subsumes Granger causality~\cite{Barnett2009}, which served for decades as
the \textit{de facto} detector of time-series
causality~\cite{granger1969investigating}. As such, transfer entropy is now
widely used in a variety of contexts including economic, biological, and
chemical processes~\cite{Bauer2007,Wibral2011}. Yet, it was recently shown
that such applications must take care to not interpret transfer entropy as
detecting information flow or causal
organization~\cite{james2016information,james2017multivariate}.
As we scale up from two random processes to the size of networks, this will be an increasingly pertinent issue. 

With respect to network systems, there is of course the classic 
discipline of network information theory which concerns itself with the
\emph{information transmission capacity} of a communication
network~\cite{el2011network}. There, given a network of rate-bound links, one
measures the aggregate rate at which multiple sources can communicate without
error to multiple receivers. One hallmark result is that for a single source
and single receiver, the capacity is determined by the \emph{max-flow min-cut
theorem} which identifies bottlenecks due to network
topology~\cite{el2011network}. This approach, though, has a rather different
focus from ours as we want to quantify information of correlated evolution
\emph{across} network layers.

%


One definition of information in a network, called \emph{graph entropy}, uses the
frequency of a node's occurrence in the orbits of the graph's automorphism
group~\cite{rashevsky1955life,Mowshowitz68}. Reference~\cite{dehmer2011history}
gives a brief history and surveys its applications. Graph entropy is easily
(and helpfully) interpreted for small graphs with substantial symmetry. It
usually generates trivial or ambiguous results on large networks, however, due
to their generic lack of perfect or near-perfect group symmetries.

As we will demonstrate,
using information measures to quantifying relationships that arise in networks, and more generally in complex systems, is advantageous for several reasons. For one,
informational measures are system-agnostic: so long as what is being studied is
well-described by random variables, it matters little over what coordinates and
with what units those variables are defined. For example, it is irrelevant if a
random variable describes fluctuations in voltage, a child's gender, or counts
of chemical species. For another, as an extension to mathematical statistics,
information measures quantify nonlinear dependencies, expanding the common
notions of correlation beyond their implied linear models. And so, information
is model independent, operating directly on the data distributions with no
assumptions as to the form of dependency. Finally, and key to our uses,
information provides the ability to compare the relative strength of correlations
across layer pairs.

\subsection{Network theory}
Approaching the analysis of multiplex networks from the network theory perspective, a natural consideration is the graph Laplacian of a network. For instance, in~\cite{braunstein2006laplacian} 
they construct a probability distribution from a network's
Laplacian. Once normalized, the Laplacian is
mathematically similar to a density matrix, the object from which the von
Neumann entropy is computed in quantum mechanics~\cite{von1955mathematical}.
Using the Laplacian for each layer in a multiplex network as a probability
distribution-like object for that layer, the similarity between layers can be quantified 
by using measures such as the Kullback-Leibler divergence between two probability distributions~\cite{Domenico15}.
Since the Laplacian expresses the ``local'' curvature about nodes in a network,
it is commonly used to analyze diffusion-like dynamics on multiplex
networks~\cite{gomez2013diffusion}.  As such, Laplacian-based analysis can capture
the similarity of diffusion processes among layers.  
However, this does not reveal layer similarities and differences that are due to complex structures to which
diffusion is insensitive.

Entropy methods have been developed to characterize multiplex network ensembles such as Ref.~\cite{bianconi2013statistical} which uses the ensemble entropy to analyze
multiplex networks with correlated layer overlaps. Yet, ensemble considerations average over the 
detailed structures needed to describe pair-wise interactions between
layers in a 
multiplex network.

In making their approaches tractable several
studies~\cite{granell2013interplay, posfai2016controllability} assumed bilayer
networks in which, by definition, only one type of coupling between layers
exists. 
Similarly, treating multiplex
networks as tensors~\cite{mucha2010community,de2013mathematical} implicitly
assumes different layers can be decomposed into linear, statistical-dependency
structures. Our empirical analyses show that each layer may have distinct
dynamical evolution 
and that there can be nonlinear relationships between layers.

Finally, Ref.~\cite{Vijayaraghavan15} introduced a multiplex Markov chain to
model the correlated evolution between different layers in a multiplex network.
The premise is that each multiplex edge in the network evolves
according to an independent and identically distributed random process. 
One can then compare the difference between when that random process uses 
a multiplex-dependent null model to a null model that assumes each layer evolves independently.
For two-layer networks this method can identify 
strong statistical correlations in structural evolution. 
Unfortunately, this method 
does not scale well with the number of layers. Nor does it allow us to
compare how strongly  layers are coupled during structural evolution.


There is some progress in developing 
information theoretic approaches to networks which rely on defining a variety of network-derived
probability distributions. Degree distributions~\cite{lin2010measuring},
deviation from mean degree distributions~\cite{sole2004information}, motif
distributions, and even configuration distributions over a network
ensemble~\cite{ji2008network} are all example probability distributions that represent
distinct aspects of a network or network ensemble. Each 
captures some specific a priori selected features. Yet, these features do not capture 
the correlated structural evolution of layers in a multiplex network. 
Here we develop a joint probability distribution for the multiplex edges in a network that will  
ultimately allow us to track 
the enhanced predictive power that
one layer can provide about the evolution of another layer.

\section{Methods}
\label{sec:method}

\subsection{Notation}
\label{subsec:notation}

We begin by introducing formal notation for defining multiplex networks.
A multiplex network is a network with many layers which share the same node set. We use calligraphic letters such as $\mathcal{U}$, $\mathcal{V}$, $\mathcal{W}$ to refer to the individual layers. The multiplex network can be represented by the graph 
$G = \left(\mathcal{N}, E^{\mathcal{U}}, E^{\mathcal{V}}, E^{\mathcal{W}}, \ldots\right)$ where $\mathcal{N}$ is the node set and $E^{\mathcal{U}}, E^{\mathcal{V}}, E^{\mathcal{W}}$ are the edge sets for each of the  different layers. $E^{\mathcal{U}}, E^{\mathcal{V}}, E^{\mathcal{W}} \subseteq {\left[\mathcal{N}\right]}^{2}$, where ${[\mathcal{N}]}^2$ denotes a Cartesian square of set $\mathcal{N}$ and, for example, $(i, j) \in E^{\mathcal{U}}$ if there is an edge between nodes $i$ and $j$ in layer $\mathcal{U}$.
In the rest of the paper, for simplicity, we consider only the case of undirected networks, so we have the additional constraint that if $(i, j) \in E^{\mathcal{U}}$ then $(j, i) \in E^{\mathcal{U}}$. However, our approach can be extended to directed networks in a straightforward manner.

\begin{figure*}[t!]
  \centering
  \includegraphics{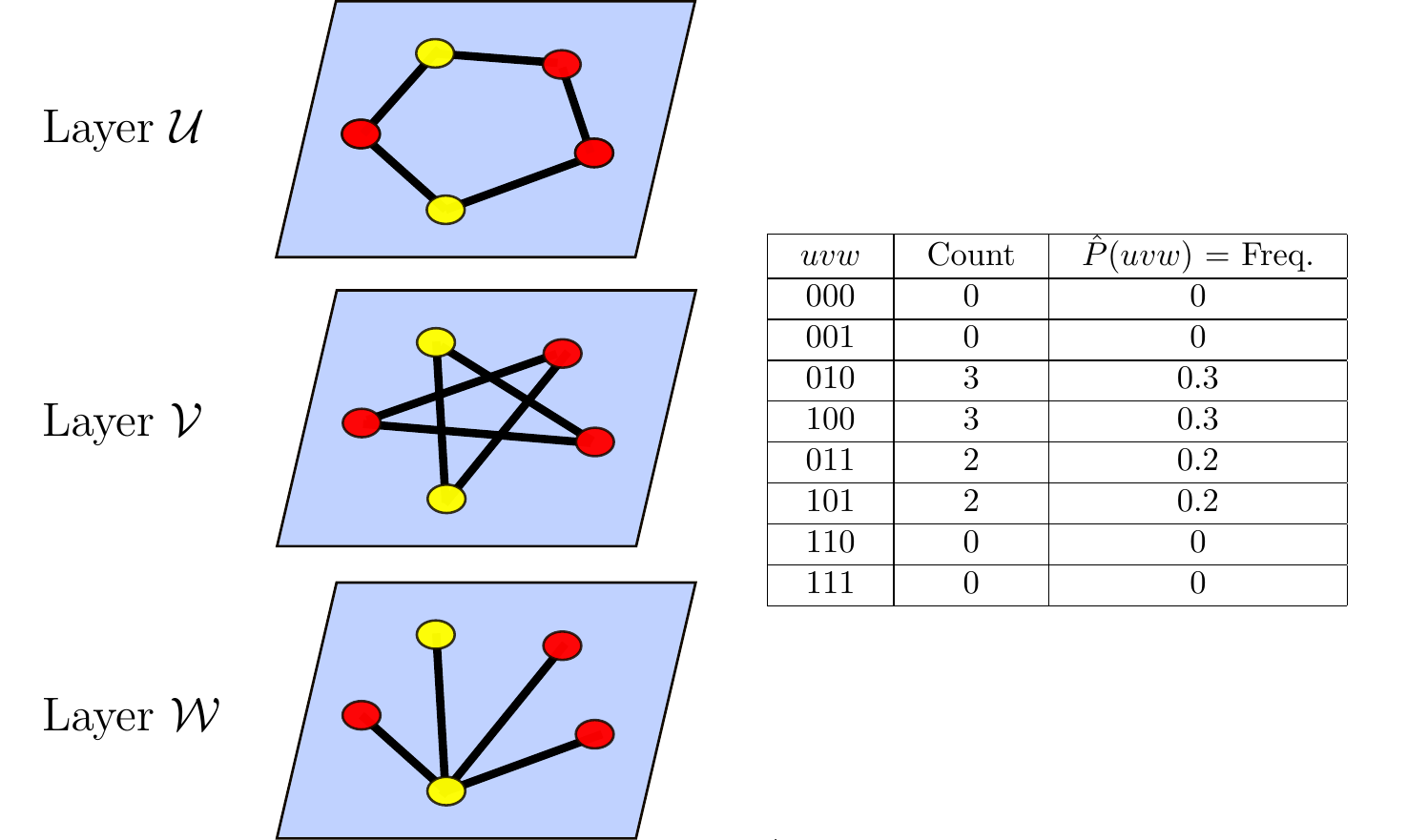}

  \caption{An example of a 3-layer multiplex network and how to build the associated joint probability distribution for the multiplex edge vectors. 
For each pair of nodes $i$, $j$, we consider the $l$-gram, $\boldsymbol{e}_{ij}$, describing the presence or absence of an edge between them respectively in each layer of the network.  
If an edge exists in a particular layer, 
we denote the corresponding element in the $l$-gram as 1, otherwise 0. We then take this edge as an instantiation of a joint probability distribution. For instance, between the two yellow nodes, there is no edge in layer $\mathcal{U}$, but there are edges in layers $\mathcal{V}$ and $\mathcal{W}$, so we have an instance of the 3-gram $\boldsymbol{e}=011$ which contributes to the tally of counts in the table row for $l$-gram $011$. We then repeat this process for all possible pairs of nodes and use the final counts to estimate the probability of having different $l$-grams. The right-hand column gives the values for these estimators of the joint probability distribution of the $l$-grams describing the multiplex edge vectors, as shown formally in Eq.~\ref{eqn-estP}.
} 
\label{fig:mi_example}
\end{figure*}

We define the {\bf multiplex edge vector} for a pair of nodes $i$ and $j$ in an $l$-layered multiplex network as $\boldsymbol{e}_{ij} = e^1_{ij}e^2_{ij} \ldots e^l_{ij}$, where each vector element $e_{ij}^\mathcal{U}=1$ if  $(i, j) \in E^{\mathcal{U}}$ and $e_{ij}^\mathcal{U}=0$ otherwise. The layers are ordered in an arbitrary but fixed manner. 
When the number of layers is small, rather than using $e_{ij}^\mathcal{U},e_{ij}^\mathcal{V}$ and $e_{ij}^\mathcal{W}$, we will use $u_{ij}$, $v_{ij}$ and $w_{ij}$, or even simply $u$, $v$ and $w$ to refer the element in the 
vector corresponding to layer $\mathcal{U}$, $\mathcal{V}$ and $\mathcal{W}$ respectively. A particular multiplex edge vector $\boldsymbol{e}_{ij}$ between nodes $i$ and $j$ can be represented as an $l$-gram, 
where $l$ is the number of layers.
An illustrative example is shown in Fig.~\ref{fig:mi_example}, where the multiplex edge vectors $\boldsymbol{e}_{ij} = e^1_{ij}e^2_{ij} \ldots e^l_{ij}$ for all possible pairs of nodes $i$ and $j$  
are enumerated for a particular example. 

In what follows we will use capital letters $U$, $V$, $W$ to refer the random variables representing corresponding layers in the joint probability distribution that will be constructed next. Thus we use different forms of the same letter depending on context. For instance, we use $\mathcal{U}$, $U$, and $u$, to refer to different concepts associated with that layer: namely the layer itself, the random variable in the joint probability distribution, and the element in a specific instantiation of a multiplex edge vector $l$-gram.

We also use some basic notation from information theory.
Following the tradition in the literature, we
use $H$ to denote Shannon entropy and $I$ to denote mutual
information.

Briefly, we are interested in the interactions between random variables.
Let $X$ denote a random variable which takes on values $x$ drawn from a
discrete set, \textit{i.e.} an alphabet $\mathcal{X}$, with probability $p(x)$.
The entropy of a random variable, $H[X]$,
is defined as:
\begin{align}
  H[X] = - \sum_{x \in \mathcal{X}}p(x) \log_2 p(x),
\end{align}
it measures the average uncertainty of random variable $X$.
$H$ is also used for defining joint entropy.
Given a set of discrete random variables $X_1, ..., X_n$ and their joint
distribution $p(x_1, ..., x_n)$,
the joint entropy $H[X_1, ..., X_n]$ is defined as
\begin{align}
  \begin{split}
  &H[X_1, ..., X_n] =\\& -\sum_{x_1 \in \mathcal{X}_1}...\sum_{x_n \in \mathcal{X}_n} p(x_1, ..., x_n) \log_2 p(x_1, ..., x_n).
  \end{split}
\end{align}

The mutual information $I[X;Y]$ between two random variables $X$ and $Y$ is defined as:
\begin{align}
  I[X;Y] = \sum_{x \in \mathcal{X}}\sum_{y \in \mathcal{Y}} p(x, y) \log_2 \frac{p(x,y)}{p(x)p(y)}k.
\end{align}
It measures how much information one random variable contains about the other.

For convenience, we also often use $H(p_1, ..., p_n)$ to denote the entropy of
a random variable with probability $p_1, ..., p_n$ for each of its possible values.
For example, $H(0.3, 0.7)$ represents the entropy of a random variable that
has two possible outcomes where the first outcome has a probability of 0.3
and the second has a probability of 0.7.

We will define more complex information measures as needed and the interested reader can refer to the classic text by Cover \& Thomas~\cite{cover2012elements} for more
information on this topic.

\subsection{The joint probability distribution of a multiplex network}
\label{subsec:characterizing}

Here we show how it is possible to characterize a multiplex network by a joint probability distribution. 
%
The overall idea is straightforward, given that all layers have the same node set.
In the classical Erd{\"o}s-R{\'e}nyi random graph model with $N$ nodes,
each edge is independently included in the network with probability $p$~\cite{Erdos1959random}.
If we only care about the existence of an arbitrary edge 
the probability of existence for that particular edge is drawn from a Bernoulli distribution
with probability $p$.
Similarly in a multiplex network with $l$ layers,
there is an analogous construction where all of the $l$ elements of a particular $\boldsymbol{e}_{ij}$ can be drawn
from an arbitrary joint probability distribution over $l$ discrete events.

First we must introduce a basic formulation of a multiplex network.
Consider the following simple model to generate a $N$-node multiplex network, starting from $N$ isolated nodes. 
Assume that for each pair of nodes $i$ and $j$, 
the multiplex edge vector $\boldsymbol{e}_{ij}$,
is formed following the same independent stochastic process.
Recall that each element of the multiplex edge vector $\boldsymbol{e}_{ij}$ 
indicates whether there is an edge or not in the corresponding layer.
Let $P(uvw\ldots)$ denote the probability that a randomly chosen $\boldsymbol{e}_{ij}$ is equal to the particular $l$-gram $uvw...$, 
where $P(u=1)$ and $P(u=0)$ are respectively the marginalized probability
that $i$ and $j$ are connected or not connected in layer $\mathcal{U}$.
Then we can generate random $\boldsymbol{e}_{ij}$'s drawn from this distribution for all $i,j$ pairs and from that 
construct a corresponding instance of an $N$-node multiplex network. 

Under these same assumptions, given a real system,
we can get an estimate of the distribution $P(uvw...)$ from the given data.
We assume that each $\boldsymbol{e}_{ij}$
follows the same distribution independently between all $i,j$ pairs, thus each multiplex edge vector observed in a real network 
can be treated as a sample for inference.
In a $N$-node multiplex network, there are $\frac{N(N-1)}{2}$ pairs of nodes, therefore there are $\frac{N(N-1)}{2}$ number of $\boldsymbol{e}_{ij}$'s.
Note, that there are only $2^l$ distinct values that the $\boldsymbol{e}_{ij}$'s can take on (since each vector element
must be either $0$ or $1$), 
and each distinct value can be written as a distinct $l$-gram.
Note that we may not see the occurrence of all possible $l$-grams in a specific real-world instance of a multiplex network.

We next will use the frequency of occurrence for each distinct $l$-gram
to construct an estimator of $P(uvw...).$
First we count the number of times a particular $l$-gram occurs, and introduce the following function to do so:
\begin{align}
  \begin{split}
  count(uvw\ldots) = \sum_{i, j \in \mathcal{N}} & \left[\left[(i, j) \in E^\mathcal{U}\right]=u\right] \\
  & \left[\left[\left(i, j\right) \in E^\mathcal{V}\right]=v\right] \\
  & \left[\left[\left(i, j\right) \in E^\mathcal{W}\right]=w\right] \\
  & \ldots
  ~,
  \end{split}
\end{align}
where we have used Iverson brackets, $[P]$, a generalization of the Kronecker delta; it evaluates to 1 if the proposition $P$ inside it is True and 0 otherwise~\cite{knuth1992two}.

With the counts in place, we can 
construct 
an estimator of the probability distribution $P(uvw...)$, explicitly: 
\begin{align}
  \begin{split}
    \hat{P}\left(U=u\ and\ V=v\ and\ W=w\ldots\right)\\
    =\frac{count(uvw\ldots)}{N(N-1)/2},
  \end{split}
  \label{eqn-estP}
\end{align}
where $N=\left|\mathcal{N}\right|$.
Once we have the estimated probability distribution, we can construct random variables and directly calculate information theoretic measures.

%

\begin{figure*}[t!]
  \centering
  \includegraphics{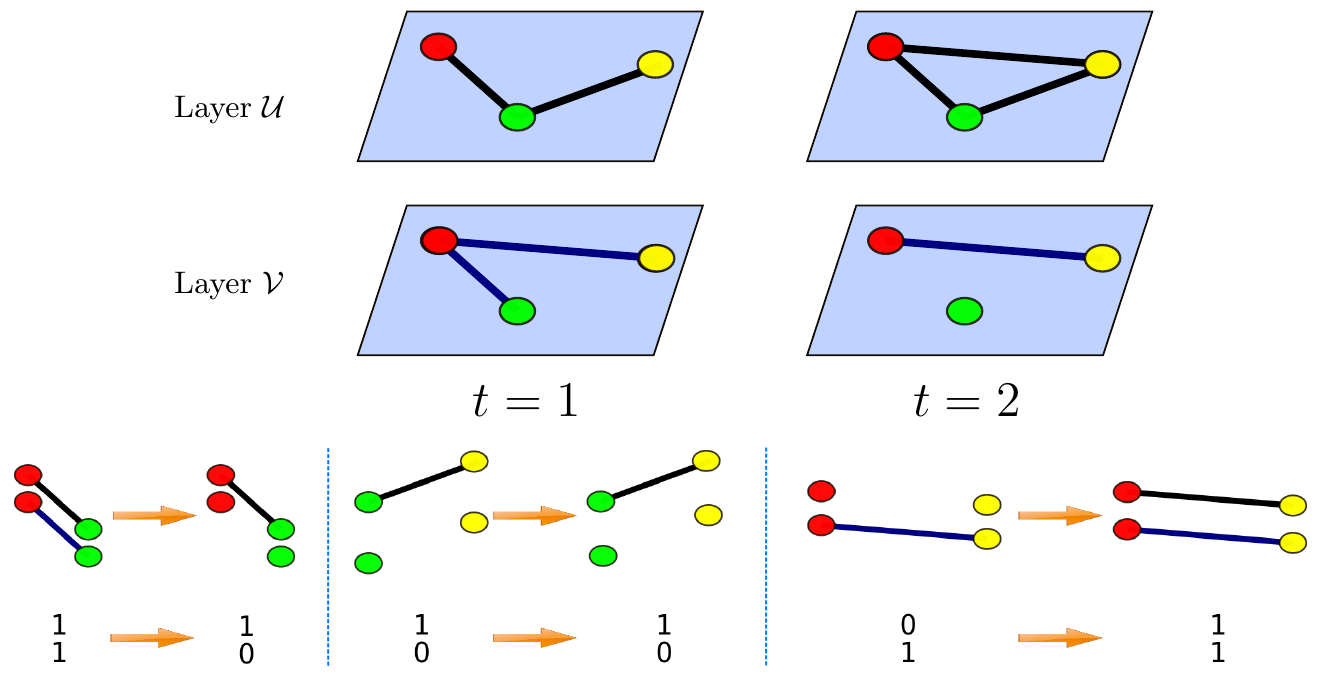}
  \caption{An example of how to construct random variables describing the discrete time evolution of a two-layer multiplex network. In this simple example there are three pairs of nodes and the evolution of each pair can be encoded by a $4$-gram describing the presence or absence of edges.  Each $4$-gram encodes a specific pattern of evolution. For instance the $4$-gram 1110 describes the evolution of the edges between the red and green nodes, while 1010 describes the evolution of the edges between the green and yellow nodes. A count of the $4$-grams observed across all pairs of nodes provides an estimator of the joint probability distribution for each possible $4$-gram.}
\label{fig:example}
\end{figure*}

Figure~\ref{fig:mi_example} shows an example of how the estimator of
a joint probability distribution can be constructed from an instance of a multiplex network with three layers $\l{U}$, $\l{V}$ and $\l{W}$. It is 
established by counting the frequency of occurrence for the different values of the multiplex edge vectors (which are $3$-grams for this example). 
Each layer of the multiplex network has a corresponding 
random variable in such a joint probability distribution which we denote by $U$, $V$ and $W$ respectively.
The random variables take on the values of either one or zero indicating, respectively, the presence or absence of an edge in the corresponding layer. 

A similar method can also be applied to analyze the discrete time dynamics of a multiplex network. For each time step $t$, $t+1$, $\ldots$, all of the layers can be included in a composite multiplex network $G=\left(\mathcal{N}, E^{\mathcal{U}^t}, E^{\mathcal{U}^{t+1}}, E^{\mathcal{V}^t}, E^{\mathcal{V}^{t+1}}, \ldots\right)$. We can then construct a time-labeled probability distribution for the network at each of these time steps. This process is demonstrated in Figure~\ref{fig:example} for a two-layered multiplex network with layers $\mathcal{U}$ and $\mathcal{V}$. For each particular pair of nodes we can denote their specific evolution over two consecutive time steps by a 
$4$-gram defined as $u^{t} v^{t} u^{t+1} v^{t+1}$ where $u^{t} = 0$ if there is no edge in layer $\l{U}$ between these two nodes in time step $t$ and $u^t=1$ if there is an edge (and respectively for $v^{t}$ and $v^{t+1}$). For example, the $4$-gram $1011$ represents the case that in time step $t$ there is an edge between these two nodes in layer $\l{U}$ but not in layer $\l{V}$ and there are edges in both layers $\l{U}$ and $\l{V}$ in time step $t+1$. By counting the frequency of these $4$-grams among all pairs of nodes in the multiplex networks, we can have an estimator of the joint probability distribution $P(u^{t} v^{t} u^{t+1} v^{t+1})$.

\subsection{Correlations between layers: \\ Mutual Information}
\label{subsec:mi}

Given the estimator of the joint probability distribution 
we can construct
information measures using that inferred distribution. 
Most important in our context is the mutual information between layers which provides us a way to quantify the extent of their correlation. This has a clean null model that all layers are statistically independent, in which case the mutual information 
between them is 0. 
Another advantage of utilizing the mutual information is that it also captures anti-correlation, the scenario where existence of edge in one layer signals the decreased likelihood of having an edge between the same pair of nodes in another layer, which is not captured by basic edge overlap considerations~\cite{Battiston14}.

For the example shown in Figure~\ref{fig:mi_example}, the mutual information
between the random variables $U$, $V$ and $W$ can be constructed in a pairwise manner.
With the standard Shannon entropy and mutual information notation~\cite{cover2012elements}
the mutual information between layer $\mathcal{U}$ and $\mathcal{V}$ is:
\begin{align}
  \begin{split}
    I\left[U; V\right] &= H\left[U\right] + H\left[V\right] - H\left[U, V\right] \\
           &= H\left(0.5, 0.5\right) + H\left(0.5, 0.5\right) \\
           & \qquad - H\left(0.0, 0.5, 0.5, 0.0\right) \\
           &= 1\ \textrm{bit}
  \end{split}
\end{align}
$H[U]$ is obtained by counting the $1$-grams present in layer $\mathcal{U}$. There we have
5 edges (``1"s) and 5 non-edges (``0"s) and therefore $H[U] = H(5/10, 5/10)$.
$H[V]$ is obtained the analogous manner.  
For $H[U,V]$, we count the $2$-grams formed by layers $\mathcal{U}$ and $\mathcal{V}$.
We then have 0 ``00" and ``11"s values, 5 ``01"s values and 5 ``10"s values.
Therefore $H[U,V] = H(0/10, 5/10, 5/10, 0/10)$.  The resulting mutual information of 1 bit is consistent with intuition: layers $\mathcal{U}$ and $\mathcal{V}$ are complementary and therefore there is maximal mutual information between them. 

We can also calculate the mutual information between layer $\mathcal{U}$ and layer $\mathcal{W}$ and between layer $\mathcal{V}$ or layer $\mathcal{W}$:
\begin{align}
  \begin{split}
    I\left[U; W\right] &= H\left[U\right] + H\left[W\right] - H\left[U, W\right] \\
           &= H\left(0.5, 0.5\right) + H\left(0.4, 0.6\right) \\
           & \qquad - H\left(0.3, 0,2, 0.3, 0.2\right) \\
           &= 0\ \textrm{bit}
  \end{split}
\end{align}

\begin{align}
  \begin{split}
    I\left[V; W\right] &= H\left[V\right] + H\left[W\right] - H\left[V, W\right] \\
           &= H\left(0.5, 0.5\right) + H\left(0.4, 0.6\right) \\
           & \qquad - H\left(0.3, 0.2, 0.3, 0.2\right) \\
           &= 0\ \textrm{bit}
  \end{split}
\end{align}
Thus, for this example, knowing whether an edge is in either layer $\mathcal{U}$ or layer $\mathcal{V}$ is not helpful for predicting the existence of the edge in layer $\mathcal{W}$ and vice versa; the pair-wise mutual information between $\mathcal{U}$ and $\mathcal{W}$ and between $\mathcal{V}$ and $\mathcal{W}$ is 0. 

Note that this is a case where even though layers $\mathcal{U}$ and $\mathcal{W}$ have overlaps, it is due to simple randomness and therefore the existence of an edge in one layer is uninformative as to the existence of the same edge in the other layer.

We next turn to the main focus of the manuscript, which is conditional mutual information.  
We do further consider mutual information 
and apply it to real data, with the details found in Appendix~\ref{app:survey}. There we show that some pairs of layers are much more correlated than others.

\subsection{Correlated structural evolution: Conditional Mutual Information}
\label{subsec:cmi}

In this section we introduce conditional mutual information and establish 
how to use this to develop information theoretic measures to quantify the correlations present in the structural evolution of multiplex networks. 

Given three random variables $X$, $X'$ and $Y$, the conditional mutual information $I\left[X'; Y | X\right]$ is defined as the relative entropy between the joint probability distribution of $X'$ and $Y$
and the product of distributions of $X'$ and $Y$ each conditioned on $X$. Formally:
\begin{align}
  I\left[X'; Y | X\right] = \sum_{\substack{x' \in \mathcal{X}' \\ y \in \mathcal{Y}}} p\left(x',y|X\right) \log \frac{p\left(x',y|X\right)}{p\left(x'|X\right)p\left(y|X\right)}~.
\label{eqn:cmi}
\end{align}
This quantifies the amount of additional information available to predict $X'$ knowing both $Y$ and $X$, beyond simply knowing $X$ alone.
This is related to the notion of 
transfer entropy discussed briefly in Sec.~\ref{sec:related_work} which is widely used in time series analysis to quantify if one time series can be used to predict another. One nicety of this measure is that if $Y$ and $X'$ are correlated to some other confounding variable $X$, conditioning on $X$ can filter out such effects.

We introduce the notion of {\bf information theoretic influence} (denoted I-INF or simply IINF) 
which is calculated by applying
conditional mutual information to the correlated structural evolution
of a multiplex network. Using the notation introduced in Sec.~\ref{subsec:notation}, 
a pair of layers at time $t$ is represented by
the random variables $U^{t}, V^{t}$. 
Information theoretic influence (IINF) from layer
$\l{U}$ to $\l{V}$ then can be defined as the mutual information between
layer $\l{U}$ at time step $t$ and layer $\l{V}$ at time step $t+1$
conditioned on layer $\l{V}$ at time step $t$, i.e. the conditional
mutual information $I[U^{t}; V^{t+1} | V^{t}]$. Formally:
\begin{align}
  IINF_{\l{U} \rightarrow \l{V}}^{t \rightarrow t+1} = I[U^{t}; V^{t+1} | V^{t}].
\label{eqn:iinf}
\end{align}
Unlike mutual information discussed in Sec.~\ref{subsec:mi}, 
IINF is asymmetric due to 
the existence of chronological order among the random variables.
In general $IINF_{\l{U} \rightarrow \l{V}}^{t \rightarrow t+1}$ is not
equal to $IINF_{\l{V} \rightarrow \l{U}}^{t \rightarrow t+1}$.
Intuitively, what $IINF_{\l{U} \rightarrow \l{V}}^{t \rightarrow t+1}$ quantifies is the amount
of extra information available to predict layer $\l{V}$ in time $t+1$ if
we also have information of layer $\l{U}$ in time $t$ in addition to information of layer $\l{V}$ at time $t$. Here conditioning on layer $\l{V}$
at time $t$ allows for the filtering out of some effects generated by
node level factors or exogenous events that happen in layer $\l{V}$. 


We then use the IINF defined in Equation~\ref{eqn:iinf}, which is essentially
a one-step \emph{transfer entropy}, to quantify the degree
to which the evolution of network layer $\l{U}$ affects layer $\l{V}$.
The intuition here is that we wish to quantify
the influence that $U^{t}$ has on $V^{t+1}$ above and beyond the influence
of $V^{t}$. This is qualitatively similar to Granger
causality~\cite{granger1969investigating}, where vector auto-regression is
used to determine if observations of a time-series $X$ improves predictions
of a time-series $Y$ beyond utilizing only observations of $Y$.
By definition, the IINF from a layer to itself
$IINF_{\l{U} \rightarrow \l{U}} = I[U^{t}; U^{t+1} | U^{t}]$
is always zero -- 
no additional information is gained through redundant knowledge of the network itself. 
Of course, the history of a particular layer may best inform the evolution of that layer, but 
in order to measure only the influence between layers we 
condition on knowledge of that particular layer. 

The Python implementations of above measures are made available~\cite{wu15multinet}, which makes use of the \texttt{dit} information
theory package~\cite{dit}.

\subsection{Illustrative Examples}

IINF measures information theoretic influence from
one layer to another. The larger the value of IINF from layer $\l{U}$ to layer $\l{V}$, the better that we can predict layer $\l{V}$ by also knowing
layer $\l{U}$'s history, beyond knowing layer $\l{V}$'s history alone.
From our construction, IINF is an information theoretic measure
applied to a binary random variable and therefore the value is always
between 0 and 1. However, dependent upon the edge density of the layers, 
the theoretical maximum is sometimes smaller than 1.
IINF 
complements
other common information theoretic measures for binary random
variables, such as entropy or mutual information.
As it is difficult to normalize IINF across different systems, it is often more useful to compare IINF between different pairs of
layers or time steps in the same multiplex network. More discussion
about possible normalization of IINF are provided in section~\ref{sec:conclusion}.

Consider the following simple cases which should serve as illustrative scenarios:
\begin{enumerate}
    \item Layer $\l{U}$ and layer $\l{V}$ are both independent Erd{\"o}s-R{\'e}nyi networks and they evolve to other independent Erd{\"o}s-R{\'e}nyi networks in the next time step. Let us denote their probability for having an edge between two nodes in each respective layer and time step as $p(\l{U}^{t})$, $p(\l{V}^{t})$ and $p(\l{V}^{t+1})$.
This represents the extreme case where two layers are totally independent and there is no information theoretic influence at all from layer $\l{U}$ to layer $\l{V}$.

In such case, the information theoretic influence is:
      \begin{align}
        \begin{split}
          IINF_{\l{U} \rightarrow \l{V}}^{t \rightarrow t+1}
          &= I\left[U^{t}; V^{t+1} | V^{t}\right]\\
          &= I\left[U^{t}; V^{t+1}\right]\\
          &= H\left[U^{t}\right] + H\left[V^{t+1}\right] - H\left[U^{t}, V^{t+1}\right] \\
          &= H\left(0, p\left(\l{U}^{t}\right)\right) + H\left(0, p\left(\l{V}^{t+1}\right)\right) \\
          &= 0\ \textrm{bit}
        \end{split}
      \end{align}
    \item Layer $\l{V}$ is an Erd{\"o}s-R{\'e}nyi network with a static structure that does not evolve in time. Then no matter what layer $\l{U}$ is, the information theoretic influence is:
      \begin{align}
        \begin{split}
          IINF_{\l{U} \rightarrow \l{V}}^{t \rightarrow t+1}
          &= I\left[U^{t}; V^{t+1} | V^{t}\right]\\
          &= H\left[U^{t} | V^{t}\right] + H\left[V^{t+1} | V^{t}\right] \\
          & \qquad - H\left[U^{t}, V^{t+1} | V^{t}\right] \\
          &= H\left[U^{t} | V^{t}\right] + 0 - H\left[U^{t} | V^{t}\right] \\
          &= 0\ \textrm{bit}
        \end{split}
      \end{align}
This represents the extreme case where layer $\l{V}$ can be perfectly predicted from itself in the previous time step and there is nothing more that we can learn from another layer no matter what.
    \item Layer $\l{U}$ and layer $\l{V}$ are both independent Erd{\"o}s-R{\'e}nyi networks and layer $\l{V}$ mimics layer $\l{U}$ in next time step, or formally $\l{V}^{t+1} = \l{U}^t$.
In this case, layer $\l{V}$ fully depends on layer $\l{U}$ in the previous time step, therefore it has maximum information theoretic influence from layer $\l{U}$ to layer $\l{V}$.

The information theoretic influence then will be:
      \begin{align}
        \begin{split}
          IINF_{\l{U} \rightarrow \l{V}}^{t \rightarrow t+1}
          &= I\left[U^{t}; V^{t+1} | V^{t}\right]\\
          &= H\left[U^{t} | V^{t}\right] + H\left[V^{t+1} | V^{t}\right] \\
          & \qquad - H\left[U^{t}, V^{t+1} | V^{t}\right] \\
          &= H\left[U^{t}\right] + H\left[U^{t}\right] - H\left[U^{t}\right] \\
          &= H\left[U^{t}\right] \\
          &= H\left(p\left(\l{U}^{t}\right)\right)
        \end{split}
      \end{align}
    \item Layer $\l{U}$ and layer $\l{V}$ are both independent Erd{\"o}s-R{\'e}nyi networks and layer $\l{V}$ is a combination of layer $\l{U}$ and layer $\l{V}$ in next time step. For the extreme case, say in the next time step the existence of an edge between two nodes in layer $\l{V}$ is the xor of the existence of the edge between the same pair of nodes in layer $\l{U}$ and layer $\l{V}$ in the previous time step, where layer $\l{V}$ is determined by a synergy effort of both layer $\l{U}$ and layer $\l{V}$.

The information theoretic influence then will be:
      \begin{align}
        \begin{split}
          IINF_{\l{U} \rightarrow \l{V}}^{t \rightarrow t+1}
          &= I\left[U^{t}; V^{t+1} | V^{t}\right]\\
          &= H\left[U^{t} | V^{t}\right] + H\left[V^{t+1} | V^{t}\right] \\
          & \qquad - H\left[U^{t}, V^{t+1} | V^{t}\right] \\
          &= H\left[U^{t}\right] + H\left[V^{t+1}\right] - H\left[U^{t}\right] \\
          &= H\left[V^{t+1}\right] \\
          &= H\left(\rho\left(\l{V}^{t+1}\right)\right)
          \end{split}
        \end{align}
\end{enumerate}
where $\rho\left(\l{V}^{t+1}\right)$ is the density of layer $\l{V}$ in time step $t+1$.

In all four extreme scenarios described above, our measure agrees with 
intuition, and for all other scenarios the IINF will fall between
these extreme cases. A few more practical examples are provided in\
Appendices~\ref{app:formation},~\ref{app:random_rewire}~and~\ref{app:merge}.

\subsection{Testing for Statistical Significance}
\label{subsec:significance}

When applied to empirical data, it is also important to be able to distinguish true signal from random fluctuations. Fortunately, methods for statistical testing of information-theoretic measures have been established and can be adopted easily.

According to Goebal \textit{et al.}~\cite{Goebel05}, in a joint probability distribution $\omega$, consider three random variables $X, Y, Z$.
If $X$ and $Y$ are independent when conditioned on $Z$,
then we can 
simply use the frequency as an estimator for probability. The inferred conditional mutual information $\hat{I}\left[X; Y | Z\right]$ by $M$ independent samples generated from $\omega$ is approximately gamma distributed:
\begin{align}
    \hat{I}\left[X; Y | Z\right] \sim \Gamma \left(\frac{\left|\mathcal{Z}\right|}{2} \left(\left|\mathcal{X}\right|-1\right)\left(\left|\mathcal{Y}\right|-1\right),\frac{1}{M \ln 2}\right)
    ~.
\end{align}
where the alphabet for these random variables are $\mathcal{X},\mathcal{Y}$ and $\mathcal{Z}$ respectively.
Applying this to network data, where the presence of an edge is binary, therefore $|\mathcal{X}| = |\mathcal{Y}| = |\mathcal{Z}| = 2$,
the gamma distribution then reduces to an exponential distribution, we find:
\begin{align}
    \hat{IINF_{\l{U} \rightarrow \l{V}}^{t \rightarrow t+1}} \sim \text{Exp} \left(\frac{1}{M \ln 2} \right)
\end{align}
where $M = N(N-1)/2$ is the number of possible edges in a network with $N$ nodes.
Then, for a given pair of layers with $\hat{IINF_{\l{U} \rightarrow \l{V}}^{t \rightarrow t+1}} = a$, the p-value would be the probability that $\hat{I}\left[U^{t}; V^{t+1} | V^{t}\right] \ge a$ if $U^{t}$ and $V^{t+1}$ are conditionally independent given $V^{t}$.
We can use this technique on real data to calculate $p$ values and establish when a correlation observed in the structural evolution is statistically significant.

\section{Applications: Correlated Structural Evolution}
\label{sec:application}

Now that we have developed a measure to quantify 
the enhanced predictive power that one layer
provides about another during structural evolution of a multiplex network,
we can show the applicability of the method. 
In this paper, we restrict our analysis to only networks evolving at consecutive time
steps in order to give a simple and clear picture. This is analogous to 
a Markov assumption, or the even weaker assumption that the most recent time step
provides the most predictive single measurement. However, it is straight
forward to extend our framework to incorporate more time steps with more
data as well. Some details about such extensions are discussed later.

To show that our method can be applied broadly 
across domains we present the results for three kinds of multiplex networks
that have intrinsically different correlation patterns between their layers:
political interactions between nation states, the commercial US airline network made of multiple carriers, and trade and alliance networks between nations.
The differences between these networks are highlighted here and details can be found later.
In the political events network, layers are different types of actions and the
correlations among the different actions 
are relatively stable over time. In
the airline network, layers represent the flight route maps of individual
airline companies within the United States and the IINF between them can change abruptly around certain
events such as mergers.
In the ally-trade network of
nation states, we have two categories of layers: one layer representing the alliances, and then many other layers representing trade of distinct types of goods. 
Correlations among two trading layers are distinct from correlations between
the alliances layer and a specific trading layer.

In all three domains, we demonstrate that IINF can quantify the correlated structural evolution
and be used to detect anomalies. 
We find that, in general, there is usually a very 
statistically significant
information theoretic influence present in the correlated structural evolution between the layers of real-world multiplex networks. 
Of course there are limitations and we will discuss them in Sec.~\ref{sec:conclusion}.

Although
we do not currently have prior knowledge about how to quantify this measure across different multiplex networks, it is very useful when applied to a specific multiplex network.  
Indeed, the IINF 
varies greatly for different pairs of layers within the
same network.
The IINF of the most strongly coupled pairs of layers can be
several orders of magnitude larger than the IINF of the majority of the rest, which
indicates a significant potential connection between how those layers choose to
create or destroy edges.
%
Moreover, we also find a correspondence between IINF
spikes and major events occurring in some of the networks,
which makes IINF a  tool for probing potential shocks to network structures.

\subsection{ICEWS Events Network}
\label{subsec:icews}

\begin{figure*}
  \centering
  \includegraphics[width=0.9\textwidth]{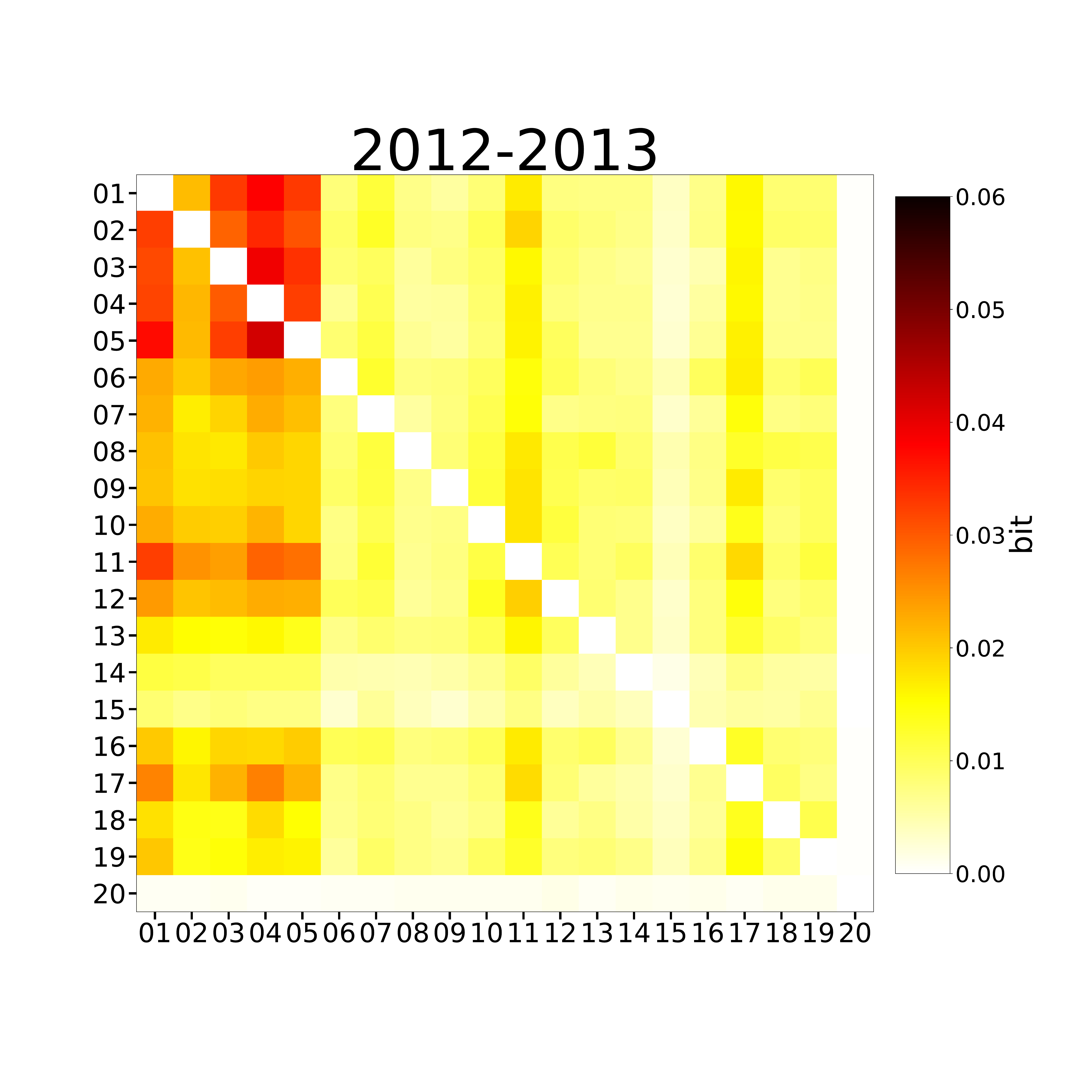}
  \caption{IINF between 20 different classes of events during the period from 2012 to 2013, with event-types labeled by their corresponding 
 CAMEO code~\cite{gerner2002conflict} (given in Appendix~\ref{app:cameo}). Each pixel represents the IINF from the event type in X-axis to the event type in Y-axis. The values that are not statistically significant from 0 in a p=0.001 level are omitted. 
 The IINF from 03 (express intent to cooperate) to 05 (engage in diplomatic cooperation) is much higher from 03 to 15 (exhibit military posture), which indicates that knowing whether countries expressed intent to cooperate in the previous time step allows us to better predict whether they actually engage in diplomatic cooperation, but knowing so will not help with better predicting actions such as military postures. We can also see that actions coded by 01 to 05, in general, provide more power for predicting other layers but that the relationship is not symmetrical. }
\label{fig:icewshm}
\end{figure*}

The Integrated Crisis Early Warning System (ICEWS) is an automatically generated dataset of international events~\cite{IcewsData}. The data contains multiple different types of political interactions between nation states ranging from making statements about one another to conducting military operations against one another. Using this data we build a series of snapshots of this multiplex network of nation states over distinct years, where each layer corresponds to a  distinct type of interaction. During the 17 year time period of our data, spanning from 1997 to 2013, the IINF
pattern is stable with just small fluctuations (See Appendix~\ref{app:stable}). We show in Fig.~\ref{fig:icewshm} the typical behavior of the pattern for the transition between two recent years. 
Thus we can use such stable patterns to help promote future predictions. This is in contrast to what we will show next in the airline networks, where IINF patterns can rapidly spike. 
The yearly transitions of a full 17-year period for the ICEWS data, in addition to the weekly transitions of a recent period, can be found in Appendices~\ref{app:stable}~and~\ref{app:timescale}.

We find that the relative correlation strength observed between layers is consistent with intuition. 
As mentioned in a previous section, the diagonal elements measure the IINF from a layer to itself
$IINF_{\l{U} \rightarrow \l{U}}$ and they are always zeros.
IINF is a directional measure and is not symmetric, which means extra information can be
more easily gained from one direction of evolution over the other.
For example, we find that the IINF from both action 03 (express intent to cooperate)
or action 04 (consult) to action 05 (engage in diplomatic cooperation) is quite high when compared to other actions, indicating that knowing whether countries expressed intent to cooperate or consult
in the previous time step allows us to better predict whether they actually engage in diplomatic cooperation in the current time step. 
In contrast, the IINF from action 03 or action 04 to action 14 (protest) or action 15
(exhibit military posture) is relatively low, indicating that knowing whether countries expressed intent to cooperate or consult
in the previous time step does not allow us to better predict the onset of actions such as protests and military posturing. 
Note that our measure only
quantifies the strength of the influence, it does not establish whether the influence is in the positive or negative direction.

Consistent with the use of information theory, IINF
can be related to how much information a source layer has and how much extra information can be possibly
gained with this knowledge in a target layer. This explains some of other features in Fig.~\ref{fig:icewshm}.
For instance, events with code 20 (use of unconventional mass violence) happen rarely and contain little
information. As such, they are 
very hard to predict from the occurrence of other actions (i.e., the row for code 20 has entries that are mostly close to 0.).
Code 20 actions also provide little information for predicting other actions (i.e., the column for code 20 has entries that are mostly close to 0.).
Actions with codes 01 to 05 contain more information than others as measured by their entropy.
They are useful for predicting other actions, such as 11, but in general are difficult to predict given other actions
(i.e., the columns for codes 01 to 05 have higher values than other columns, but the rows 
do not.).

\subsection{US Airline Network}
\label{subsec:air}

The USA Department of Transportation, Bureau of Transportation Statistics maintains a public database of the monthly report it receives from all certified USA air carriers~\cite{AirData}. Every domestic flight segment is recorded therein. For our purposes, we focus on the ``scheduled passenger service'' flights as this is representative of the air carriers' regular flight network structure. We do not include flights such as ``non-scheduled passenger service'' flights and flights with no passengers as they occasional and ad hoc and 
do not seem to reflect a carrier's network-building strategy.

We find that the spikes observed in IINF often are important signals, 
revealing the interactions between different layers. 
To demonstrate such a result, we first show that when a relatively high spike in IINF is observed between two layers, we often find co-occurring real world events associated with this spike.
Conversely, we also provide evidence that when an expected event with high impact happens, such as a merger between carriers, we see a spike in the IINF measure.

\begin{figure*}
  \includegraphics[width=0.85\textwidth]{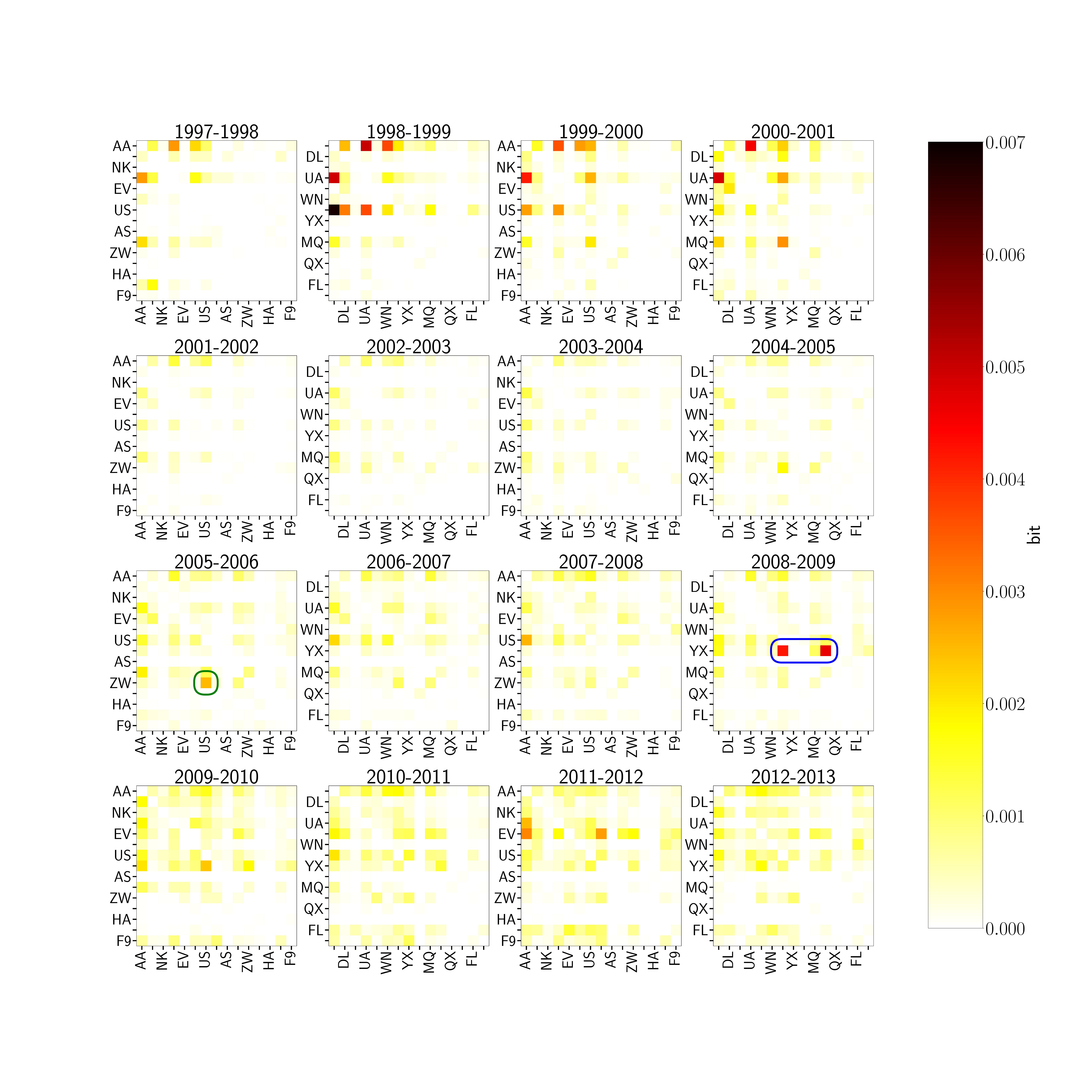}
  \caption{IINF between 15 major carriers from 1997 to 2013. (The IATA codes are indicated for every other carrier in each figure, the full list from top to bottom and also left to right are: AA, DL, NK, UA, EV, WN, US, YX, AS, MQ, ZW, QX, HA, FL, F9.) Each distinct panel 
 is the IINF between two distinct consecutive years, with each pixel representing the strength of IINF from the carrier in X-axis to the carrier in Y-axis. Values that are not statistically significant from 0 at the p=0.001 level are omitted. In general, we can see that the interactions among carriers decreased significantly after the 9/11 attacks. 
 After that, there are a few cases with unusual spikes in IINF. These are generally explainable by large events. For example, in 2009 (blue circle), Midwest Airlines (YX) is acquired by Republic Airways. The latter inherited the same IATA code from the former. They then adjusted the flight routes to compete with US Airways (US) and Air Wisconsin (ZW). Also in 2005 (green circle), Air Wisconsin (ZW) invested heavily into US Airways (US) and signed a long term contract operating as US Airways Express.} 
\label{fig:yearly}
\end{figure*}

Figure~\ref{fig:yearly} shows how IINF behaves between the 15 major airline companies during a 17 year period. In general, there are statistically significant information flows between all carrier pairs.
However, when comparing this to the IINF values present in the ICEWS events network, the magnitude of the values in the airline network are generally much lower, which suggests that the amount of influence between layers is much smaller in airline networks then in the ICEWS network. Another notable observation is that a transition happened during 2001 which could be related to the September 11 attacks (also referred to as 9/11)~\cite{ito2005assessing}. The information theoretic influence was higher before the attacks and fell-off dramatically after it. This suggests that 
heavy regulation after 9/11 may have had a significant impact, preventing carriers from adjusting their route map relative to other carriers.

\begin{figure*}
  \centering
  \includegraphics[width=0.4\textwidth]{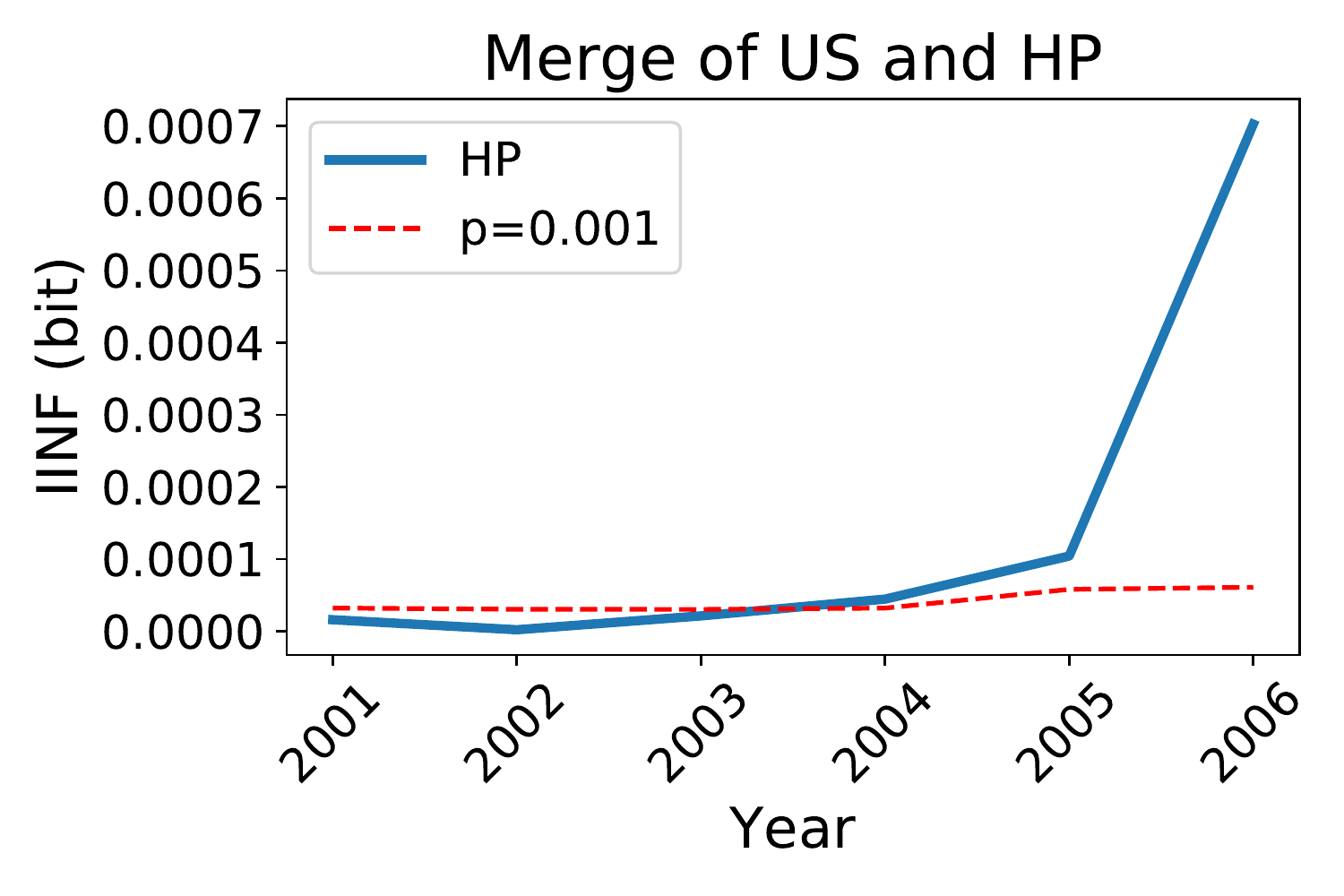}
  \includegraphics[width=0.4\textwidth]{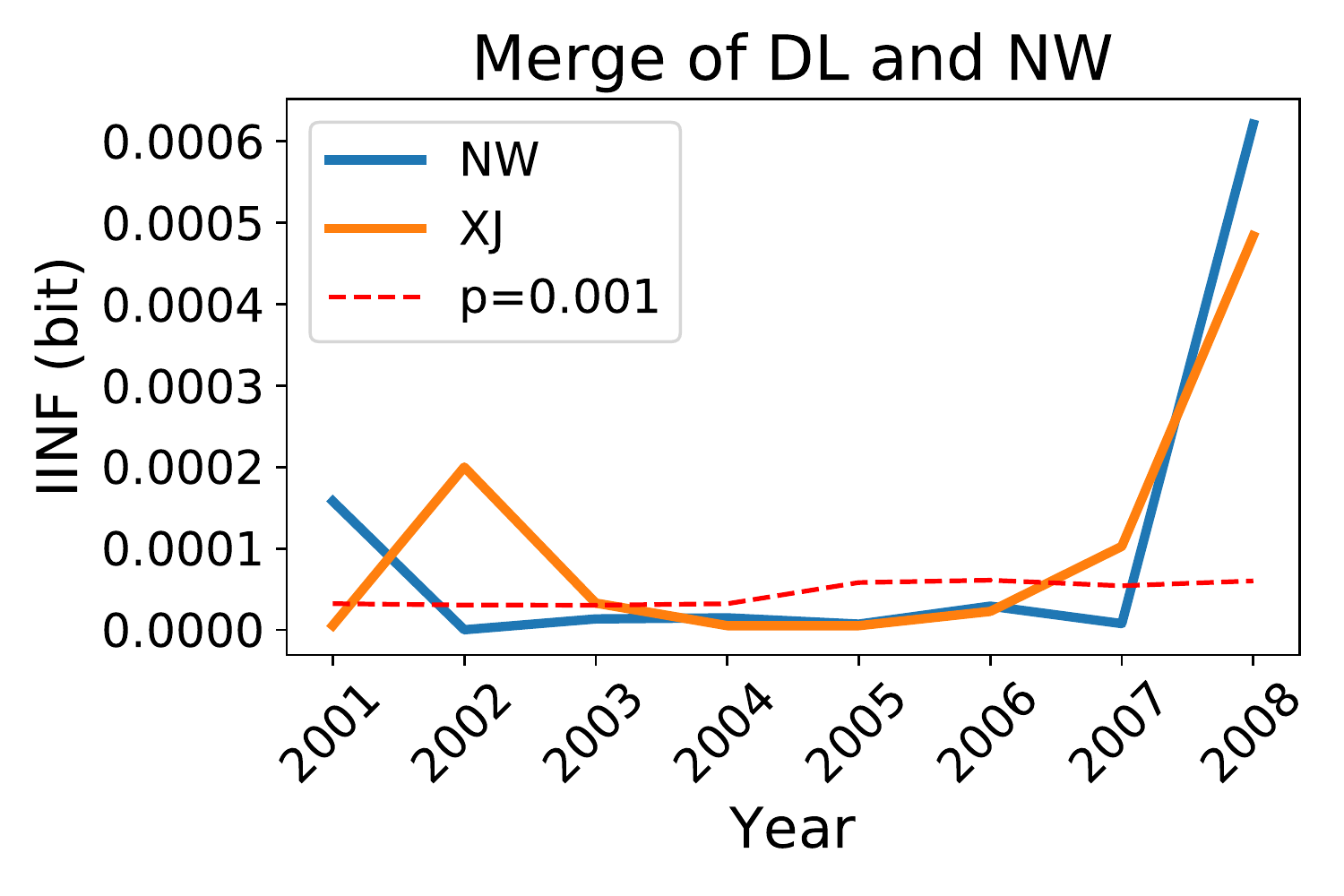}
  \includegraphics[width=0.4\textwidth]{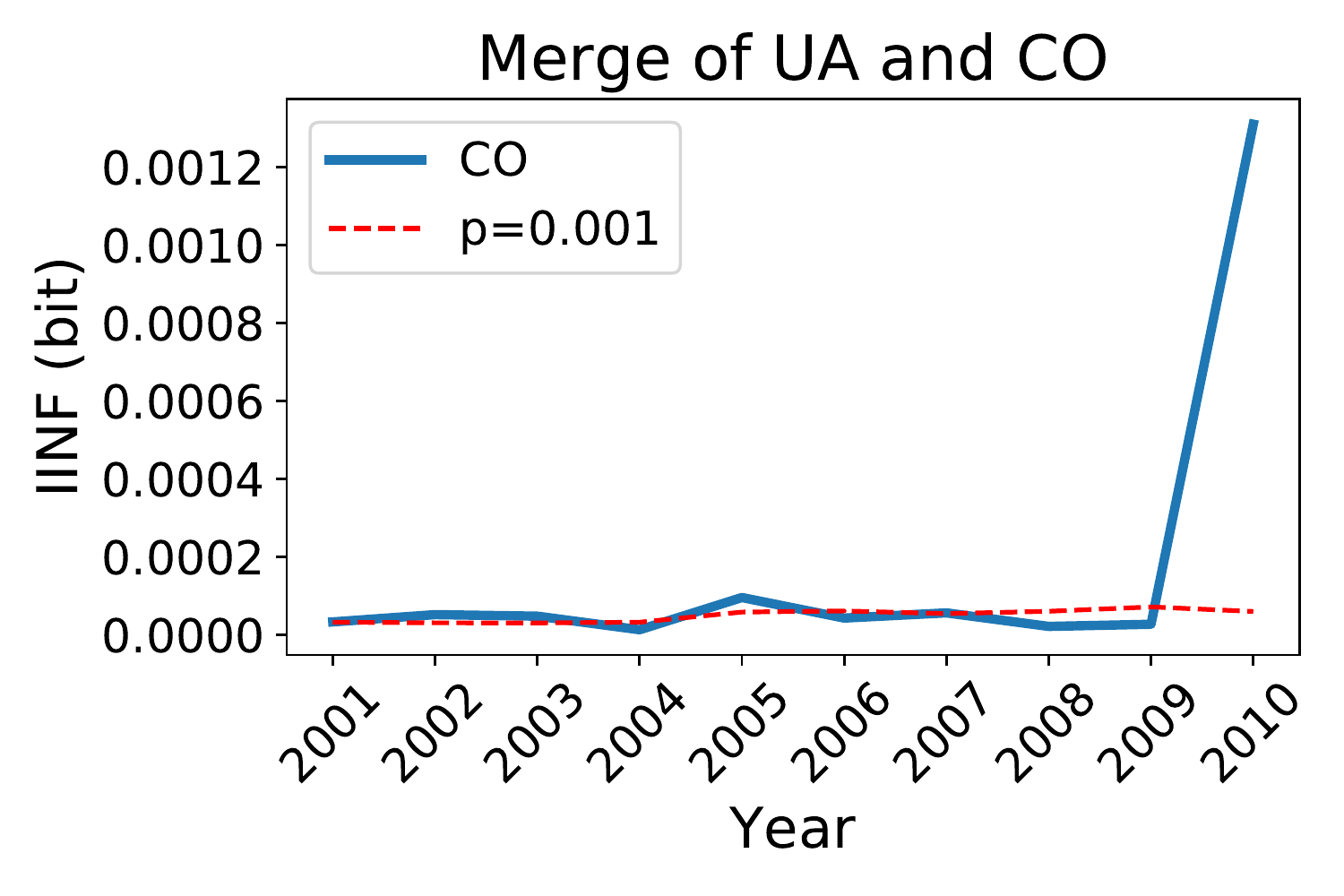}
  \caption{Changes in IINF during three different large merger events between air carriers 
 showing a spike in IINF as the carriers merge. (Note that carriers are required to continue reporting separately for one more year beyond the official merger date.) 
 American West Airlines (HP) merged with US Airways (US) in 2005. Northwest Airlines (NW) merged with Delta Air Lines (DL) in 2008 (The additional green solid line is Mesaba Airlines (XJ) who was operating routes for NW). Continental Airlines (CO) merged with United Airlines (UA) in 2010. The statistical significance level for IINF is also included.}
\label{fig:merge}
\end{figure*}

We now demonstrate a correspondence between spikes in IINF and significant real-world events. We manually identified the top three, post-9/11, IINF hot-spots and corroborated that   
each one corresponds to some associated event, including an acquisition and the signing of a long-term cooperation contract, as explained in the caption of Fig.~\ref{fig:yearly}. This is consistent with our expectation that 
carriers adjust their flight routes to take into account the routes of the carriers that they acquire or sign major contracts with.
When it is know in advance that mergers are underway, it is expected that the merging carriers 
will adjust their flight networks accordingly, and we hypothesize that this will result in an increase in IINF. Note that after a merger, carriers are still required to separately report their flight information for one additional year which allows us to corroborate this hypothesis with our dataset. Figure~\ref{fig:merge} shows details for 
three different airline carrier mergers.

We also provide a heatmap of IINF among all 60 air carriers in recent years in Appendix~\ref{app:airline}. It is interesting to note that IINF between major carriers is generally larger than among smaller carriers and with more frequent spikes.

\subsection{Alliance and Trade Network}
\label{subsec:trade}

\begin{figure*}
  \centering
  (a) \includegraphics[width=0.4\linewidth]{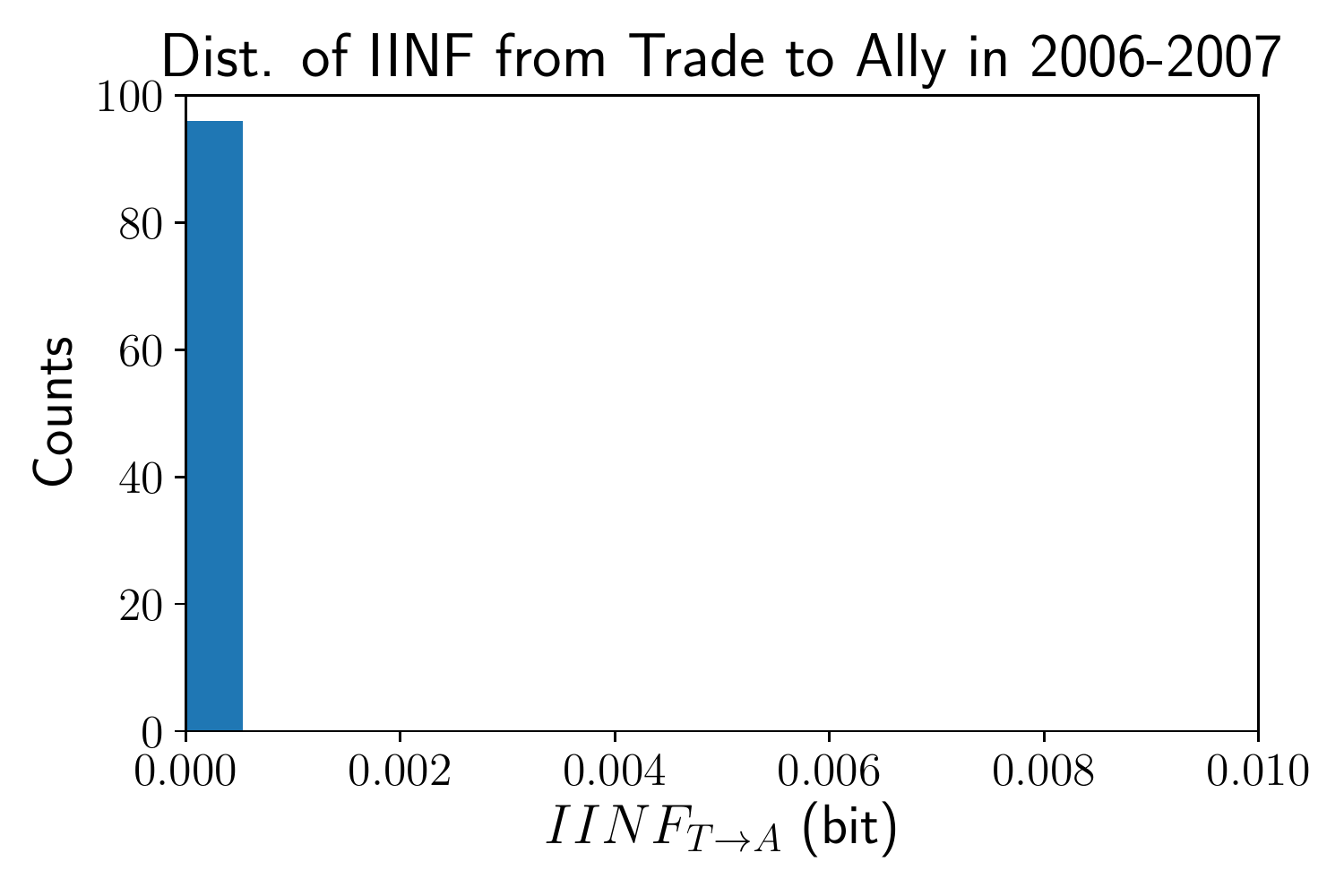}
  (b) \includegraphics[width=0.4\linewidth]{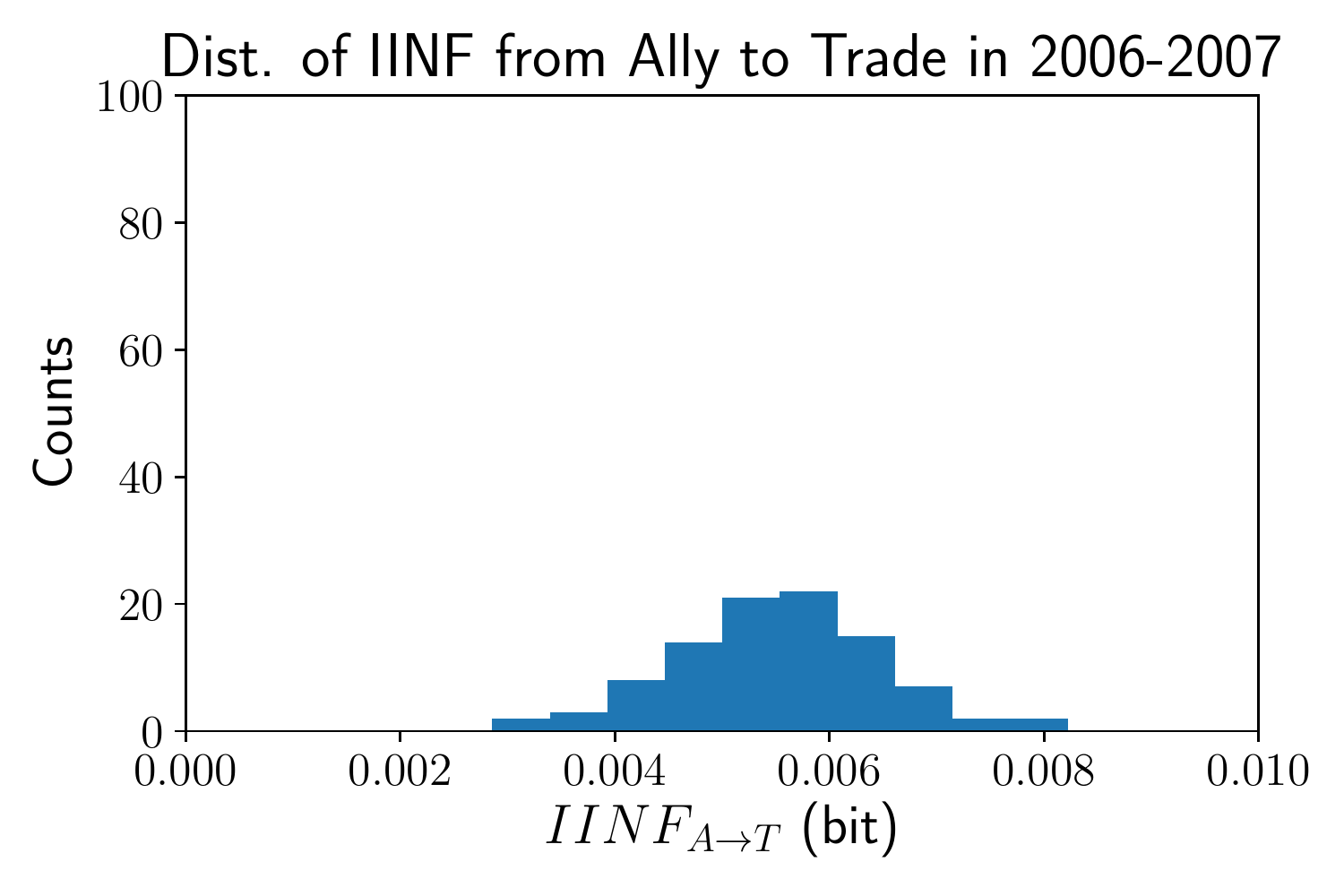}
  \caption{(a) The IINF from all the trade layers to the alliance layer is zero. (b) 
  The values of IINF from the alliance layer to the trade layers are normally distributed around non-zero values, which indicates the information flows from alliance to  trade relationships are unidirectional and somewhat ubiquitous. This could be due to the relative stability of the alliance network in the short term and quantifies prior arguments based on political and economic reasoning.}
\label{fig:allytrade}
\end{figure*}

To study the alliance and trade network between nations, we combined two different datasets. The trade network is compiled from the publicly accessible COMTRADE data maintained by the United Nations~\cite{UNComtrade}. This dataset includes yearly trade information for many different categories of goods which are hierarchically classified into a 6-digit system. For example, code 260111 represents ``Iron ore, concentrate, not iron pyrites, unagglomerate'', 2601XX represents ``Iron ores" and 26XXXX represents ``general ores and concentrates". In this research we limited ourselves to an aggregation to the first two digits to get a denser and more reliable network, which results in a 96-layer network where each layer is a distinct trade category of commodities.

The alliance network is generated from the Alliance Treaty Obligations and Provisions Project~\cite{leeds2002alliance},  containing the alliance treaties signed by nation states. We manually matched these two datasets to construct a multiplex network with one alliance layer and 96 trade layers and then studied the IINF between all the distinct alliance and trade layer pairs.

We find that for this discrete time formulation, the IINF is unidirectional, 
from the alliance network to the trade network, corroborating prior research establishing this fact from political and economic considerations~\cite{mansfield1997alliances}. Figure~\ref{fig:allytrade} shows that at this yearly time scale there is no significant IINF from any commodity trade network to the alliance network, 
but for different commodities there is typically some information that can be gained from knowledge of the alliance network in the previous year. This means that in the short term we can use the alliance network to help predict the change in the trade network but not vice versa. We also notice that the information flows from the alliance network to trade networks are small and there is no statistically significant difference 
when comparing IINF values to different commodity layers. Thus we can say, at least at this level of aggregation of the commodity categories, each category receives roughly the same IINF from the alliance network.

\section{Conclusion}
\label{sec:conclusion}

We showed that it is possible to use the edge set of a multiplex network to construct a joint probability distribution characterizing the network. Information theoretic measures over this probability distribution enable us to quantify correlations between pairs of layers, including temporal considerations. To specifically capture the extent of correlation present in the structural evolution between pairs of layers in a multiplex networks we introduce a measure called the information theoretic influence (IINF) which is based on their conditional mutual information. 
Applying this to several empirical datasets, we find that the extent of information sharing between different pairs of layers can vary dramatically in real-world multiplex networks.
%
In real-world multiplex networks, some set of layers can evolve in a highly correlated manner while other layers can evolve independently, especially when the number of layers is large. 

In addition, we show that IINF also detects asymmetric relationships between layers. For instance, political scientists hypothesize that for short term considerations the influence between trade and alliance network is unidirectional: that the alliance network drives the trade network, but that there is significantly less influence the other direction~\cite{mansfield1997alliances}. Our IINF measure quantifies this phenomena showing that, conditioned on the previous time step, a trade network provides no information for predicting the alliance network in the next step, but the alliance network does provide information for the evolution of trade networks.

Furthermore, our approach of mapping a multiplex network onto a joint probability distribution allows for 
many other information measures to be calculated. One potential direction is to use the newly developed autonomy of three-way mutual information, related to synergy and redundancy of information, to divide three-way mutual information into two types of factors~\cite{williams2010nonnegative}. One might be able to identify different signatures for different types of correlations present, such as cooperative or competitive. 
Likewise, one may be able to use these three-way measures to identify the higher-order organization in a system, beyond the dyadic organization inherent to treating a system as a network. 

We have used IINF to understand the structure and structural evolution internal to several real-world networks, but we do not currently use IINF to compare different networks. This would require understanding the proper way to normalize IINF across networks of different sizes and types of probability distributions. 
There are many ways to normalize the results so that different perspectives can be brought into consideration. For instance, 
the IINF $I\left[U^{t}; V^{t+1} | V^{t}\right]$ can be normalized by either the entropy $H\left[V^{t+1}\right]$ or the conditional entropy$H\left[V^{t+1} | V^{t}\right]$. These, respectively, would consider 
the ratio of IINF to the maximum information that can actually be obtained from the data with the layer $\l{V}$ in time step $t+1$ itself, or the layer $\l{V}$ in time step $t+1$ conditioned on layer $\l{V}$ in time step $t$. As our intent here is to use IINF within an individual multiplex network, we leave normalization considerations for future work. 

Of course there are many ways to refine the considerations introduced herein.  
For example, the assumption that all the multiplex edge vectors are drawn from a same joint probability distribution is 
not always valid.
Similarly, the weaknesses~\cite{james2016information,james2017multivariate} of the conditional mutual information should be addressed head on in future efforts.
However, we believe this provides a useful framework for quantifying correlations present between layers in a multiplex network including in their co-evolution.  

\vspace{0.1in}\noindent
{\bf Acknowledgments:} We thank Brandon Kinne and Martin Hilbert for many
useful discussions and for providing us with the ICEWS and COMTRADE datasets.
We are grateful to a number of others from the SPINS, complexnets, and
Complexity Sciences Center at UC Davis who provided insightful discussions and
feedback. We acknowledge support from the U.S. Army Research Office under
Multidisciplinary University Research Initiative Award No. W911NF-13-1-0340
and Cooperative Agreement No. W911NF-09-2-0053, the U.S. Department of Defense
Minerva Grant No. W911NF-15-1-00502, and Intel Corporation through the Intel
Parallel Computing Center at the UC Davis Complexity Sciences Center.

\appendix

\section{CAMEO code for ICEWS data}
\label{app:cameo}

Table~\ref{tab:CAMEO} gives the codebook for different layers classified with CAMEO code~\cite{gerner2002conflict} in the ICEWS data. Each layer corresponds to a different type of interaction between nation states. Refer to the codebook cited for more details.

\definecolor{lightgray}{gray}{0.9}
\begin{table}
  \centering
  \footnotesize
  \begin{tabular*}{0.5\textwidth}{rl}
	\cline{1-2}
    \textbf{Code} & \textbf{Meaning} \\
	\cline{1-2}\\[-0.2cm]
\rowcolor{lightgray}
    01 & MAKE PUBLIC STATEMENT \\
    02 & APPEAL \\
\rowcolor{lightgray}
    03 & EXPRESS INTENT TO COOPERATE \\
    04 & CONSULT \\
\rowcolor{lightgray}
    05 & ENGAGE IN DIPLOMATIC COOPERATION \\
    06 & ENGAGE IN MATERIAL COOPERATION \\
\rowcolor{lightgray}
    07 & PROVIDE AID \\
    08 & YIELD \\
\rowcolor{lightgray}
    09 & INVESTIGATE \\
    10 & DEMAND \\
\rowcolor{lightgray}
    11 & DISAPPROVE \\
    12 & REJECT \\
\rowcolor{lightgray}
    13 & THREATEN \\
    14 & PROTEST \\
\rowcolor{lightgray}
    15 & EXHIBIT FORCE POSTURE \\
    16 & REDUCE RELATIONS \\
\rowcolor{lightgray}
    17 & COERCE \\
    18 & ASSAULT \\
\rowcolor{lightgray}
    19 & FIGHT \\
    20 & USE UNCONVENTIONAL MASS VIOLENCE \\
	\cline{1-2}
  \end{tabular*}
\caption{CAMEO codes used in the ICEWS dataset.}
\label{tab:CAMEO}
\end{table}

Table~\ref{tab:CAMEO_example} is a sample of the explanation taken from the codebook
for a subcategory of 01 (MAKE PUBLIC STATEMENT).

\begin{table}
  \centering
  \begin{tabular}{lp{0.32\textwidth}}
    \textbf{CAMEO} & \textbf{011} \\
    \cline{1-2}
    \textbf{Name} & \textbf{Decline comment} \\
    \textbf{Description} & Explicitly decline or refuse to comment on a situation.\\
    \textbf{Usage Notes} & This event form is a verbal act. The target could be who the source actor declines to make a comment to or about.\\
    \textbf{Example } & NATO on Monday declined to comment on an estimate that Yugoslav army and special police troops in Kosovo were losing 90 to 100 dead per day in NATO air strikes.\\
    \cline{1-2}
  \end{tabular}
  \caption{Sample CAMEO code from codebook.}
  \label{tab:CAMEO_example}
\end{table}

\section{Mutual information in real multiplex networks}
\label{app:survey}

Mutual information explained in section~\ref{subsec:mi} can be used to quantify the correlation between two layers in a multiplex network when defined as described in the main text. It provides a principled way of quantifying the relationships between layers without assumptions such as linear correlation and it frees us from consulting a null model to verify that the correlation is not from sheer randomness. With the maturity of information theory, we can also easily obtain many statistical tools to test the significance of the results.

\begin{figure*}
    \centering
    \includegraphics[width=\textwidth]{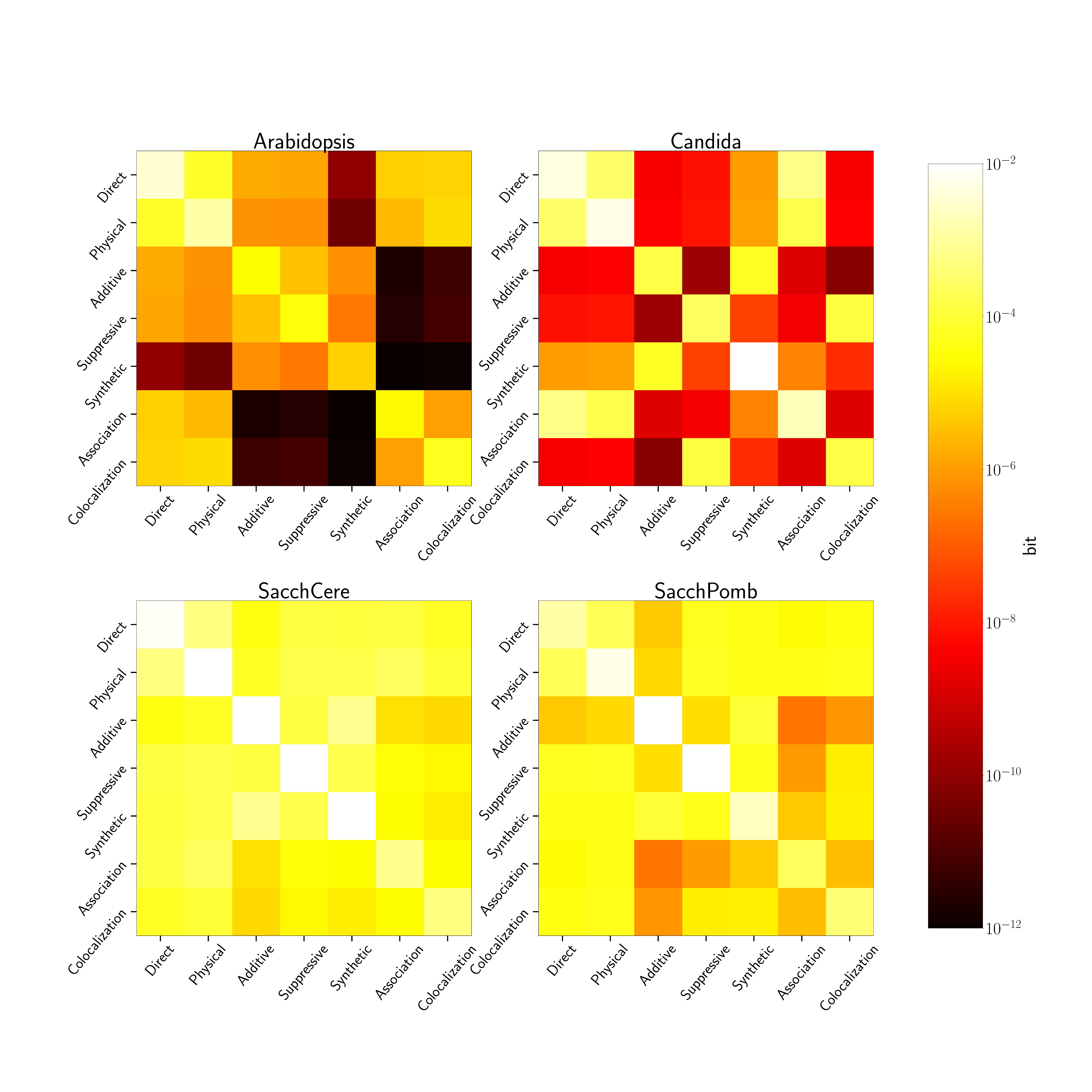}
    \caption{Mutual Information between layers in four different biology networks. The elements on the diagonal can be seen as layers' entropy.}
    \label{fig:biohm}
\end{figure*}

\begin{figure*}
    \centering
    \includegraphics[width=\textwidth]{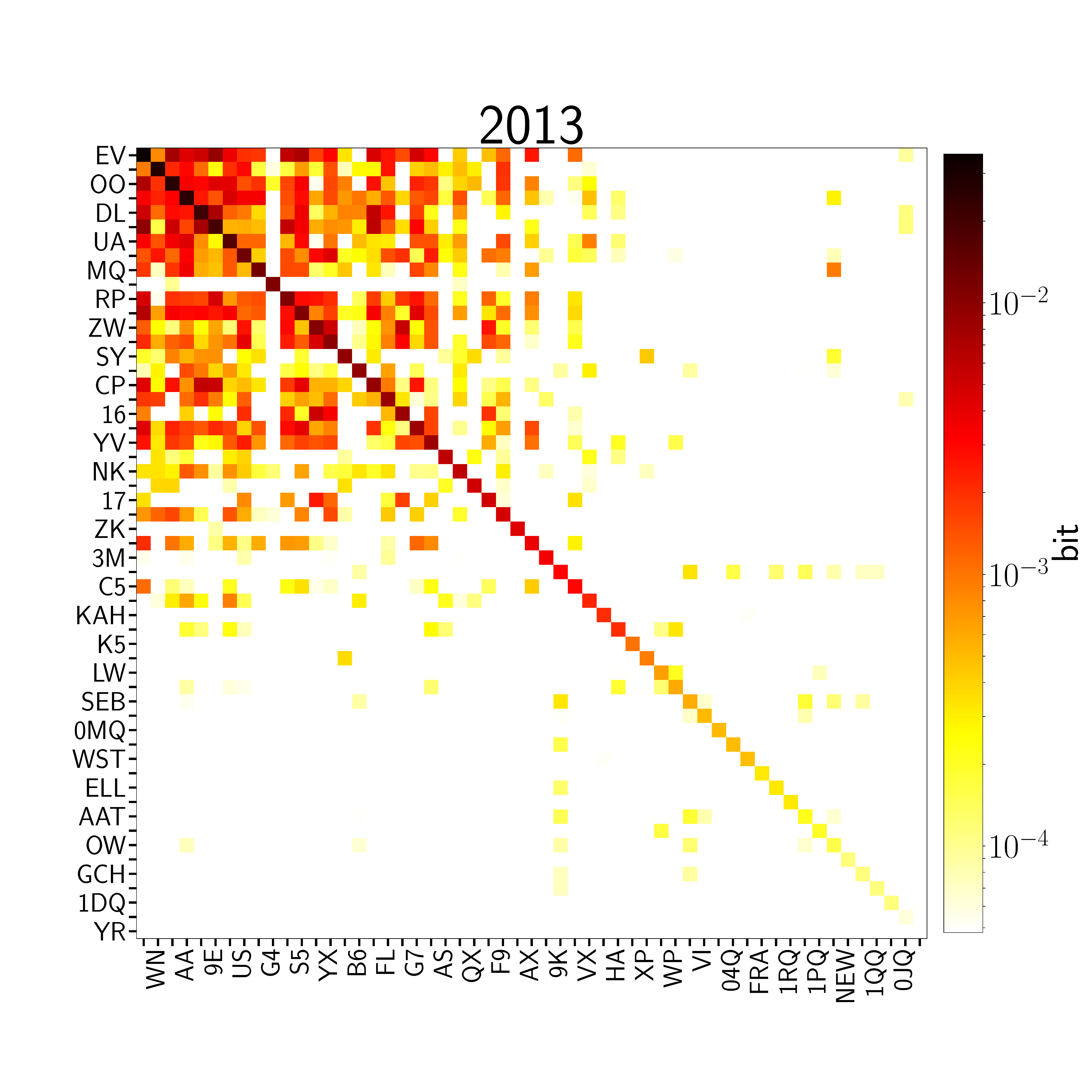}
    \caption{Mutual Information between airline carriers in 2013. Carriers are ordered from left to right and top to bottom
      and every other carriers are labeled.
      The values that are not statistically significant from 0 in a p=0.001 level are truncated. The elements on the diagonal can be seen as layers' entropy.}
    \label{fig:airhm}
\end{figure*}

\begin{figure*}
    \centering
    \includegraphics[width=\textwidth]{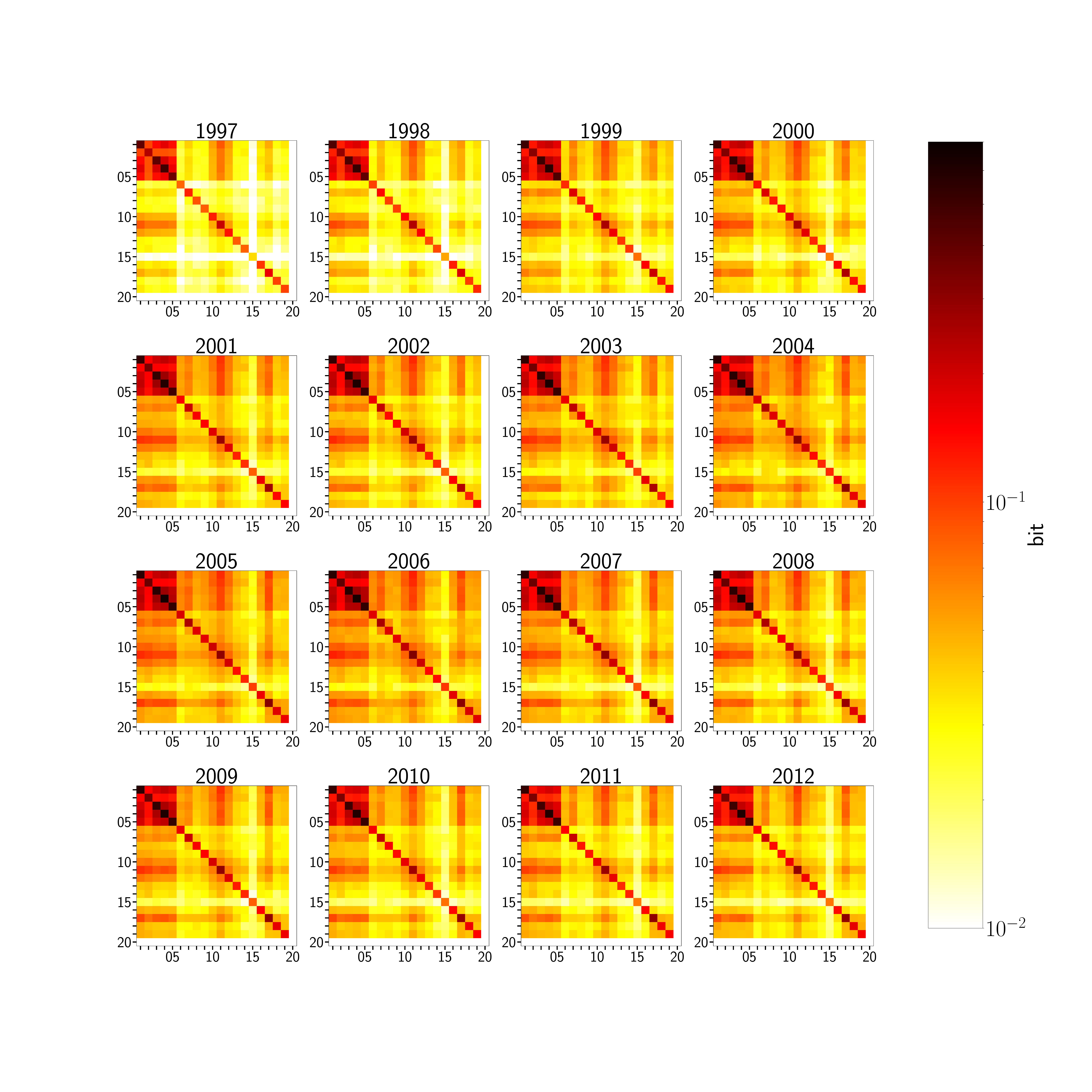}
    \caption{Mutual Information between layers in ICEWS events network from 1997 to 2012. Types of interactions are labeled by their CAMEO codes from 01 to 20, ordered from left to right and top to bottom. The elements on the diagonal can be seen as layers' entropy.}
    \label{fig:miicews}
\end{figure*}

We apply this measure to different data sets and report here in figures~\ref{fig:biohm},~\ref{fig:airhm},~and~\ref{fig:miicews}.
Figure~\ref{fig:biohm} shows the mutual information for four biological interaction
network built from the BioGrid dataset~\cite{stark2006biogrid}. The multiplex networks
represent different types of interactions between proteins/genes in four different species.
Figure~\ref{fig:airhm}~and~\ref{fig:miicews} give the mutual information between layers
in the Transtat and ICEWS datasets described in section~\ref{subsec:air}~and~\ref{subsec:icews}
respectively.
Note that mutual information between layers is symmetric:
\begin{align}
  I[U;V] = I[V;U].
\end{align}
The diagonal elements are mutual information between one layer
and itself,
\begin{align}
  I[U;U] = H[U]
\end{align}
therefore can been seen as the entropy of the
layer and all other mutual information are strictly less
than the entropy.
As mentioned in Section~\ref{subsec:mi}, these results show a great variety of correlation
strength between different layers. Note that these figures use log scales and there could
often be several orders of magnitudes difference between different layers.

\section{Stable Patterns in ICEWS data}
\label{app:stable}

\begin{figure*}
  \centering
  \includegraphics[width=\textwidth]{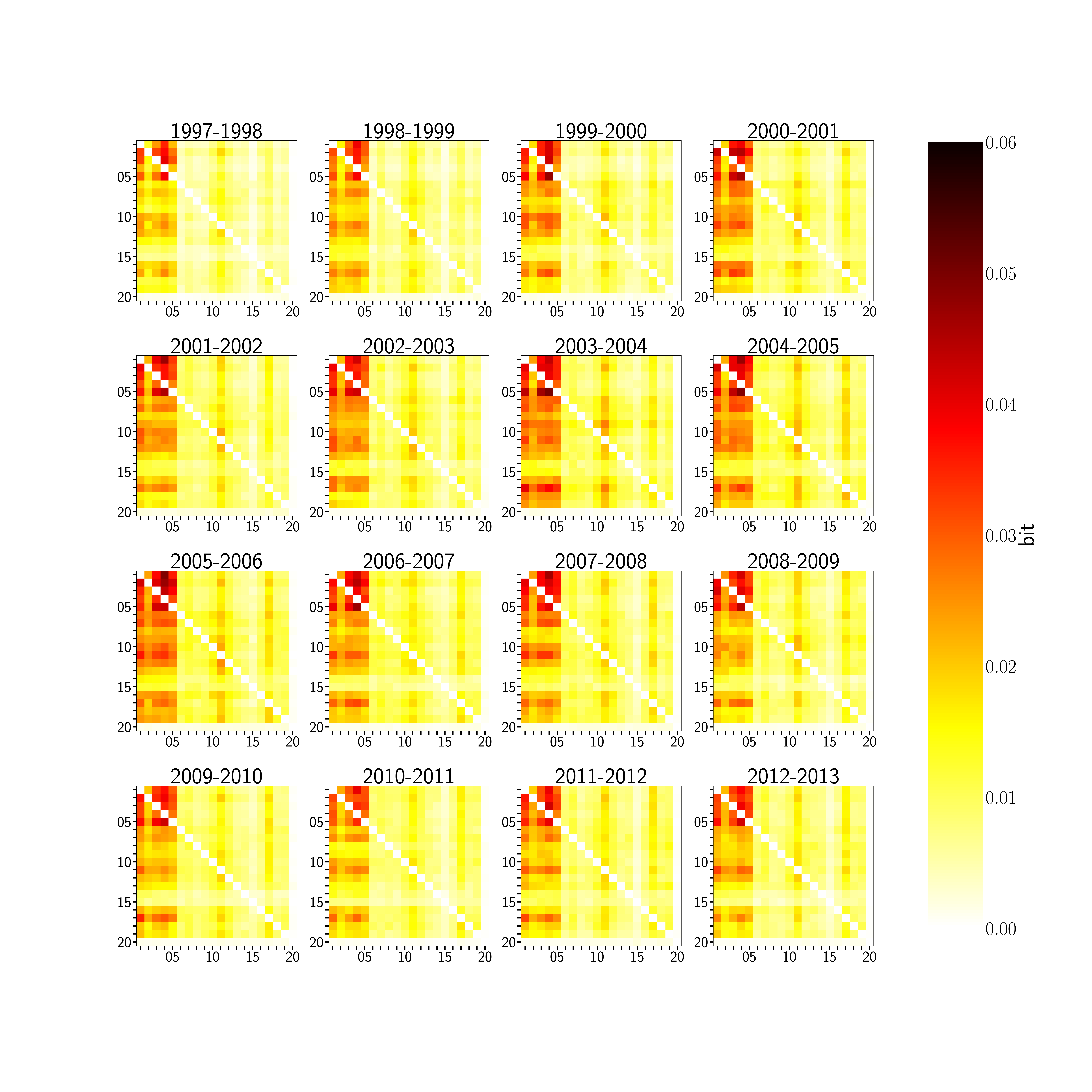}
  \caption{IINF between layers in ICEWS events network from 1997 to 2013. Types of interactions are labeled by their CAMEO codes from 01 to 20, ordered from left to right and top to bottom. The elements on the diagonal are all 0 by definition.}
\label{fig:cmiicews}
\end{figure*}

In section~\ref{subsec:icews} we show a typical pattern of the IINF in ICEWS data, here we provides a 17-year period of IINF in ICEWS data demonstrating the patterns of IINF among different years remain to be similar. This is indicating that the underlying mechanism of how different types of interactions affect each other holds constant over the studied time period. In contrast, we see that in airline networks the influence of one carrier upon another changes over time which reflects the changing interaction among carriers over different years. This can be seen in Fig.~\ref{fig:cmiicews}.

\section{Time Scale and Aggregation}
\label{app:timescale}

\begin{figure*}
  \centering
  \includegraphics[width=\textwidth]{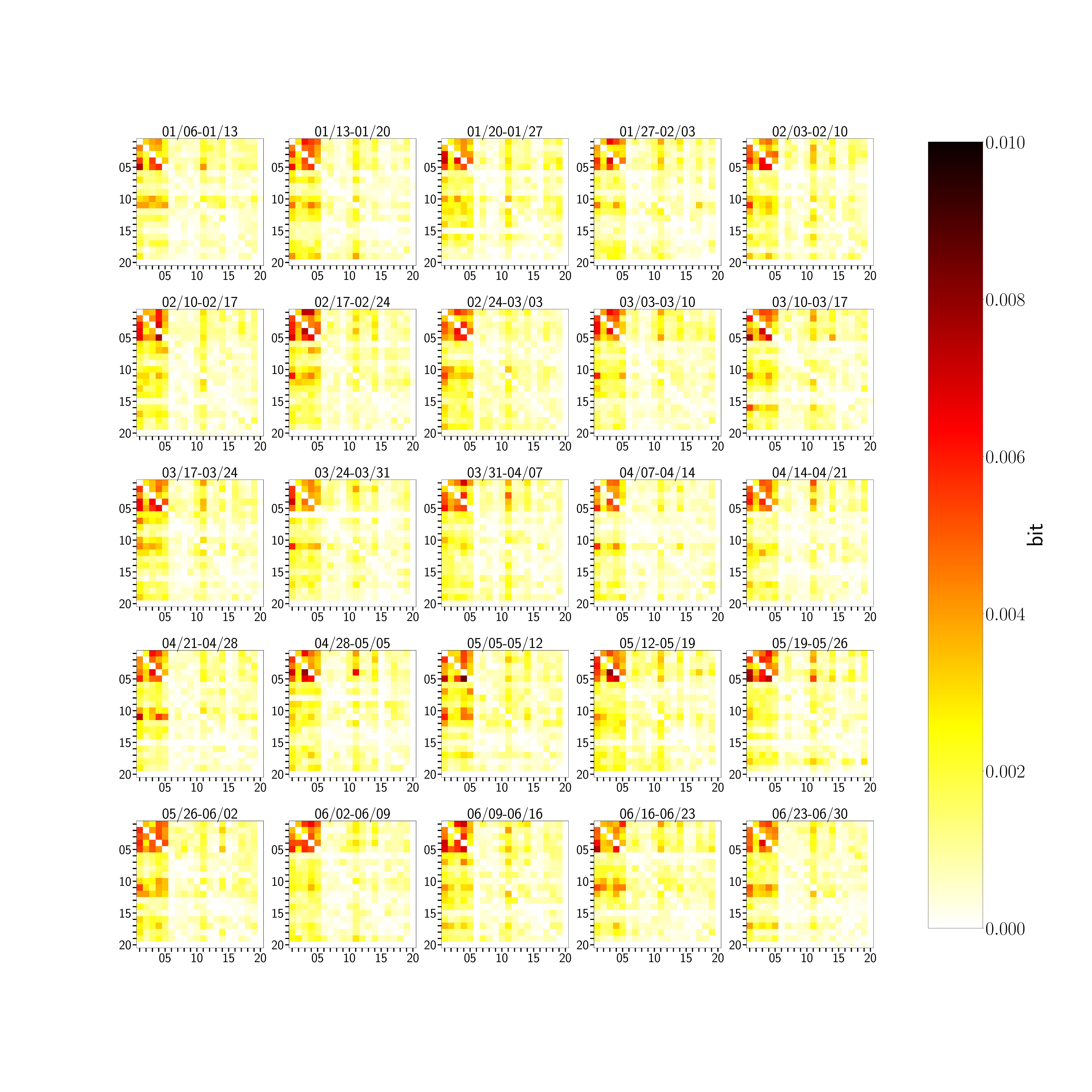}
  \caption{IINF between layers in ICEWS events network during first 26 weeks of 2014. Types of interactions are labeled by their CAMEO codes from 01 to 20, ordered from left to right and top to bottom. The elements on the diagonal are all 0 by definition.}
\label{fig:hmweek}
\end{figure*}

\begin{figure*}
  \centering
  \includegraphics[width=\textwidth]{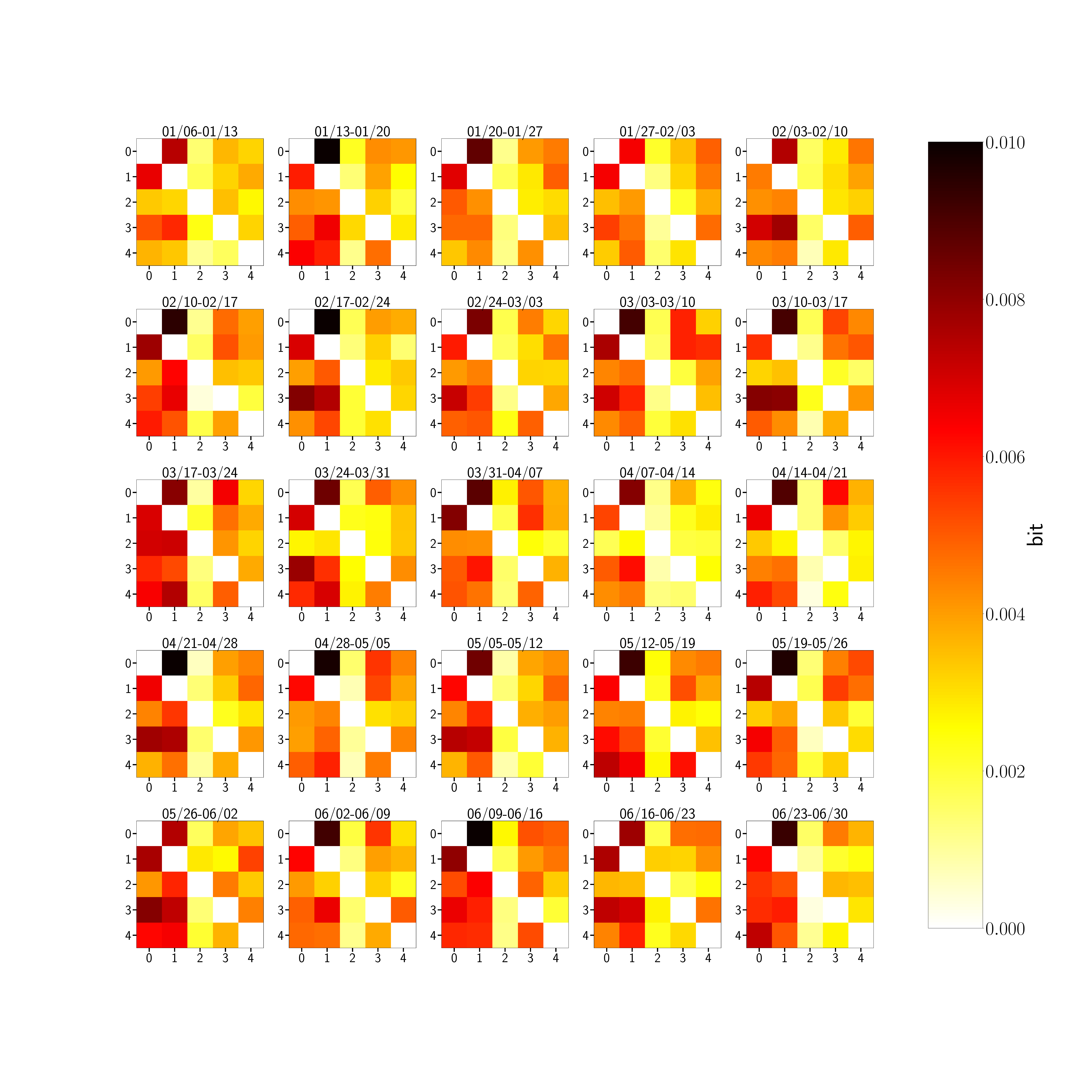}
  \caption{IINF between layers in ICEWS events network during first 26 weeks of 2014. Aggregated to penta class~\cite{duval1980reconsidering}. The elements on the diagonal are all 0 by definition.}
\label{fig:hmpenta}
\end{figure*}

When applying the information theoretic influence to quantifying the correlated structural evolution of multiplex networks, one must also be aware that this measure is sensitive to the choice of time step like transfer entropy. Here we provide a weekly IINF of ICEWS data for comparison in Figures~\ref{fig:hmweek}~and~\ref{fig:hmpenta}.

The result from weekly snapshots is qualitatively similar to what we have from yearly snapshots.
The patterns remain stable with slight variations.
Quantitatively, the magnitude of weekly IINF is smaller which indicate a weaker influence between layers.
This is suggesting that for a shorter time period,
how one types of interaction helps predicting another
is qualitatively similar to a longer time period,
but the predicting power is weaker and more volatile.

Another notable factor is how the layer is constructed from the data. Often times there are many ways to interpret data into a multiplex network, the correlated structural evolution according to those different constructions are likely to behave differently as well. As an example, ICEWS data can also be classified into a multiplex network using events' penta class, which is a higher level aggregation of CAMEO code~\cite{duval1980reconsidering}. The results with this method is also presented here. This specific observation also shows a potential direction we could pursue. By minimizing the information theoretic influence between layers, we might be able to divide a multiplex network into a few relatively independent components and study them independently.

\section{IINF of airline network}
\label{app:airline}

\begin{figure*}
  \centering
  \includegraphics[width=\textwidth]{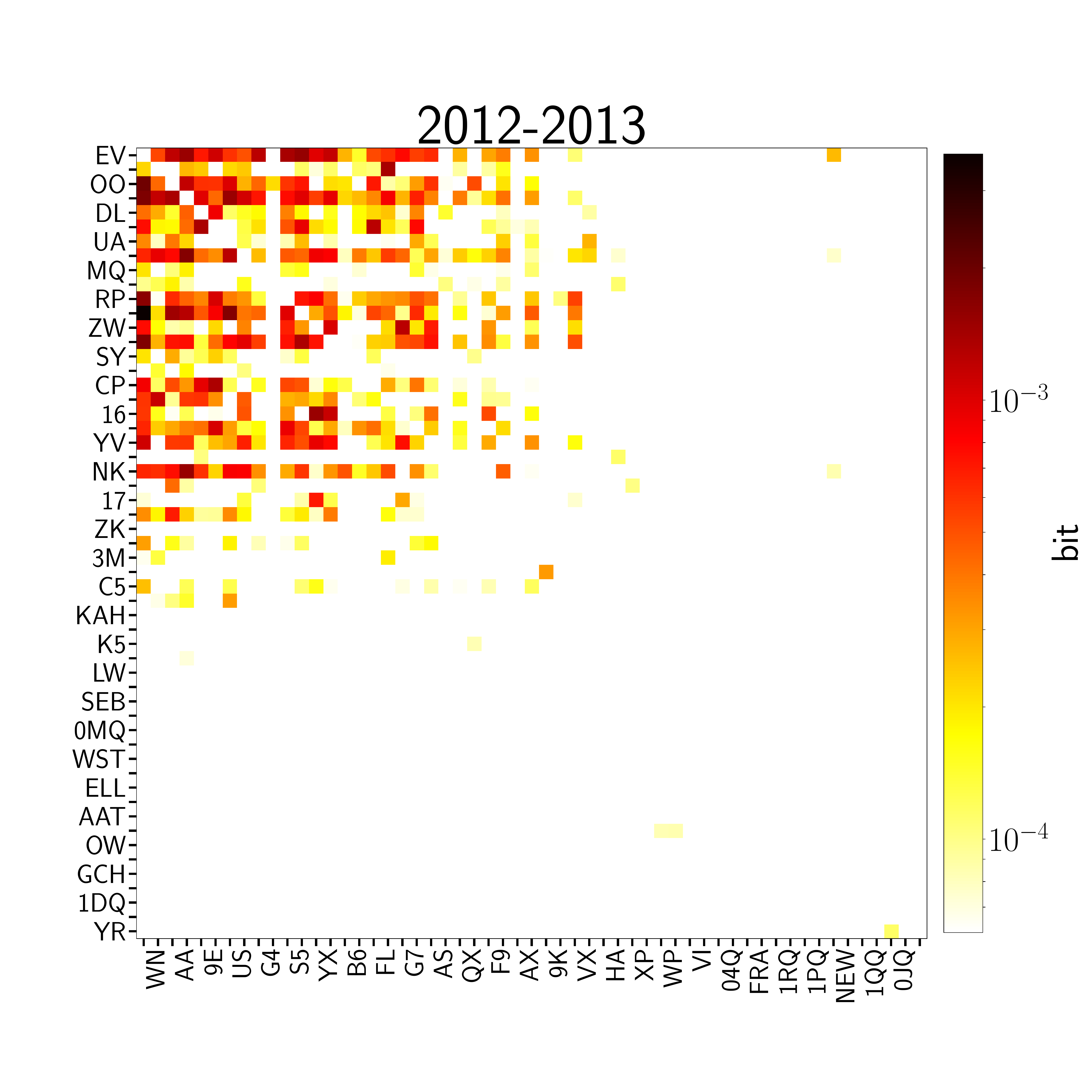}
  \caption{IINF between all pairs of carriers in 2012-2013 in log scale. Carriers are ordered from left to right and top to bottom
    and every other carriers are labeled. The values that are not statistically significant from 0 in a p=0.001 level are truncated.
    Here we can see that the information theoretic influence are mainly between those major carriers. Smaller carriers are generally neither influential nor influenced by others. The elements on the diagonal are all 0 by definition.}
\label{fig:airhm_b}
\end{figure*}

See Figure~\ref{fig:airhm_b} for the information theoretic influence among airlines between the years 2012 and 2013.
There are two notable points as mentioned in section~\ref{subsec:air}:
The IINF varies greatly among layers and most of influence happens between major carriers,
suggesting that some carriers influence each other much more and those influence happen
more frequently among major carriers that may caused by their cooperation or competition.

\section{MI and IINF in network formation}
\label{app:formation}

At the first glance, it may not be clear whether this framework also works for network formation models beyond Erd{\"o}s-R{\'e}nyi, but a carefully thought could show that it is not unreasonable to apply it to others. Let's consider a configuration model where two layers are independently generated from two arbitrary degree sequences. For every pairs of nodes $i$, $j$, the probability that there is an edge between $i$ and $j$ in layer $U$ is $p\left(U_{ij}\right)$. The estimated mutual information and information theoretic influence through our method then will still be 0 since $p\left(U_{ij} | V_{ij}\right) = p\left(U_{ij}\right)$ for all the edges.

\section{MI and IINF for random rewiring}
\label{app:random_rewire}

Consider a network with $N$ nodes and $M$ edges, let $\Sigma = N(N-1)/2$ and $\rho = M/\Sigma$. If we randomly rewire $k$ edges in such a network, the mutual information between the network before rewiring and after rewiring is then:
\begin{align}
I\left(G ; G'\right) = H\left(\rho\right) - \left[\rho H\left(\frac{k}{M}\right) + \left(1-\rho\right) H\left(\frac{k}{\Sigma-M}\right)\right]
\end{align}
  
Suppose instead we have two networks $G$ and $H$ both with $N$ nodes and $M$ edges and rewire $k$ edges of $G$ results $G'$ that is the same as $H$, then the information theoretic influence is:
\begin{align}
  \begin{split}
    IINF_{H \rightarrow G}
    &= I\left(H : G' | G\right)\\
    &= \rho H\left(\frac{k}{M}\right) + \left(1-\rho\right) H\left(\frac{k}{\Sigma-M}\right)
  \end{split}
\end{align}

\section{IINF during merge}
\label{app:merge}

Consider two networks $G$ and $H$ with $N$ nodes and $M_1$ and $M_2$ edges respectively. Suppose the number of overlapped edges are $k$ and $G'$ is the simple aggregation between $G$ and $H$, then
we have:
\begin{equation}
  \begin{split}
    IINF_{H \rightarrow G}
    &= I\left(H : G' | G\right)\\
    &= H\left(\frac{M_2-k}{\Sigma-M_1}\right)\\
    &-\frac{M_1}{\Sigma} \left[H\left(\frac{M_2-k}{\Sigma-M_1}\right) +
  H\left(\frac{k}{M_1}\right)\right]
  \end{split}
\end{equation}

\bibliographystyle{unsrt}
\bibliography{multiinfo}

\begin{thebibliography}{10}

\bibitem{Newman10}
M.~Newman.
\newblock {\em {Networks: an introduction}}.
\newblock Oxford university press, 2010.

\bibitem{Zanin15}
M.~Zanin.
\newblock Can we neglect the multi-layer structure of functional networks?
\newblock {\em {Physica A: Statistical Mechanics and its Applications}},
  430:184--192, 2015.

\bibitem{Kivela2014}
M.~Kivel{\"a}, A.~Arenas, M.~Barthelemy, J.~P. Gleeson, Y.~Moreno, and M.~A.
  Porter.
\newblock Multilayer networks.
\newblock {\em Journal of Complex Networks}, 2(3):203--271, 2014.

\bibitem{Boccaletti2014}
S.~Boccaletti, G.~Bianconi, R.~Criado, C.~I. Del~Genio,
  J.~G{\'o}mez-Garde{\~n}es, M.~Romance, I.~Sendi{\~n}a-Nadal, Z.~Wang, and
  M.~Zanin.
\newblock The structure and dynamics of multilayer networks.
\newblock {\em Physics Reports}, 544(1):1--122, 2014.

\bibitem{bianconi2018multilayer}
G~Bianconi.
\newblock {\em Multilayer networks: structure and function}.
\newblock Oxford university press, 2018.

\bibitem{Battiston14}
F.~Battiston, V.~Nicosia, and V.~Latora.
\newblock Structural measures for multiplex networks.
\newblock {\em Phys. Rev. E}, 89:032804, Mar 2014.

\bibitem{mansfield1997alliances}
E.~D. Mansfield and R.~Bronson.
\newblock Alliances, preferential trading arrangements, and international
  trade.
\newblock {\em American Political Science Review}, 91(1):94--107, 1997.

\bibitem{kantz2004nonlinear}
H.~Kantz and T.~Schreiber.
\newblock {\em Nonlinear time series analysis}, volume~7.
\newblock Cambridge university press, 2004.

\bibitem{james2016information}
R.~G. James, N.~Barnett, and J.~P. Crutchfield.
\newblock Information flows? {A} critique of transfer entropies.
\newblock {\em Phys. Rev. Lett.}, 116(23):238701, 2016.

\bibitem{james2017multivariate}
R.~G. James and J.~P. Crutchfield.
\newblock Multivariate dependence beyond shannon information.
\newblock {\em Entropy}, 19(10):531, 2017.

\bibitem{Schreiber00}
T.~Schreiber.
\newblock Measuring information transfer.
\newblock {\em Phys. Rev. Lett.}, 85:461--464, Jul 2000.

\bibitem{Barnett2009}
L.~Barnett, A.~B. Barrett, and A.~K. Seth.
\newblock Granger causality and transfer entropy are equivalent for {Gaussian}
  variables.
\newblock {\em Phys. Rev. Lett.}, 103(23):238701, 2009.

\bibitem{granger1969investigating}
C.~W.~J. Granger.
\newblock Investigating causal relations by econometric models and
  cross-spectral methods.
\newblock {\em Econometrica: Journal of the Econometric Society}, pages
  424--438, 1969.

\bibitem{Bauer2007}
M.~Bauer, J.~W. Cox, M.~H. Caveness, J.~J. Downs, and N.~F. Thornhill.
\newblock Finding the direction of disturbance propagation in a chemical
  process using transfer entropy.
\newblock {\em IEEE Transactions on Control Systems Technology}, 15(1):12--21,
  2007.

\bibitem{Wibral2011}
M.~Wibral, B.~Rahm, M.~Rieder, M.~Lindner, R.~Vicente, and J.~Kaiser.
\newblock Transfer entropy in magnetoencephalographic data: Quantifying
  information flow in cortical and cerebellar networks.
\newblock {\em Progress in Biophysics and Molecular Biology}, 105(1):80--97,
  2011.

\bibitem{el2011network}
A.~El~Gamal and Y.-H. Kim.
\newblock {\em Network information theory}.
\newblock Cambridge University Press, 2011.

\bibitem{rashevsky1955life}
N.~Rashevsky.
\newblock Life, information theory, and topology.
\newblock {\em The Bulletin of Mathematical Biophysics}, 17(3):229--235, 1955.

\bibitem{Mowshowitz68}
A.~Mowshowitz.
\newblock Entropy and the complexity of graphs: {I}. {An} index of the relative
  complexity of a graph.
\newblock {\em Bulletin of Mathematical Biophysics}, 30(1):175--204, 1968.

\bibitem{dehmer2011history}
M.~Dehmer and A.~Mowshowitz.
\newblock A history of graph entropy measures.
\newblock {\em Information Sciences}, 181(1):57--78, 2011.

\bibitem{braunstein2006laplacian}
S.~L. Braunstein, S.~Ghosh, and S.~Severini.
\newblock The laplacian of a graph as a density matrix: a basic combinatorial
  approach to separability of mixed states.
\newblock {\em Annals of Combinatorics}, 10(3):291--317, 2006.

\bibitem{von1955mathematical}
J.~Von~Neumann.
\newblock {\em Mathematical foundations of quantum mechanics}.
\newblock Number~2. Princeton University Press, Princeton, New Jersey, 1955.

\bibitem{Domenico15}
M.~De~Domenico, V.~Nicosia, A.~Arenas, and V.~Latora.
\newblock Structural reducibility of multilayer networks.
\newblock {\em Nature Communications}, 6, 2015.

\bibitem{gomez2013diffusion}
S.~Gomez, A.~Diaz-Guilera, J.~Gomez-Gardenes, C.~J. Perez-Vicente, Y.~Moreno,
  and A.~Arenas.
\newblock Diffusion dynamics on multiplex networks.
\newblock {\em Physical Review Letters}, 110(2):028701, 2013.

\bibitem{bianconi2013statistical}
G.~Bianconi.
\newblock Statistical mechanics of multiplex networks: Entropy and overlap.
\newblock {\em Physical Review E}, 87(6):062806, 2013.

\bibitem{granell2013interplay}
C.~Granell, S.~G\'omez, and A.~Arenas.
\newblock Dynamical interplay between awareness and epidemic spreading in
  multiplex networks.
\newblock {\em Phys. Rev. Lett.}, 111:128701, Sep 2013.

\bibitem{posfai2016controllability}
M.~P{\'o}sfai, J.~Gao, S.~P. Cornelius, A.~Barab{\'a}si, and R.~M. D'Souza.
\newblock Controllability of multiplex, multi-time-scale networks.
\newblock {\em Physical Review E}, 94(3):032316, 2016.

\bibitem{mucha2010community}
P.~J. Mucha, T.~Richardson, K.~Macon, M.~A. Porter, and J.~Onnela.
\newblock Community structure in time-dependent, multiscale, and multiplex
  networks.
\newblock {\em Science}, 328(5980):876--878, 2010.

\bibitem{de2013mathematical}
M.~De~Domenico, A.~Sol{\'e}-Ribalta, E.~Cozzo, M.~Kivel{\"a}, Y.~Moreno, M.~A.
  Porter, S.~G{\'o}mez, and A.~Arenas.
\newblock Mathematical formulation of multilayer networks.
\newblock {\em Physical Review X}, 3(4):041022, 2013.

\bibitem{Vijayaraghavan15}
V.~S. Vijayaraghavan, P.-A. No{\"e}l, Z.~Maoz, and R.~M. D’Souza.
\newblock Quantifying dynamical spillover in co-evolving multiplex networks.
\newblock {\em Scientific Reports}, 5, 2015.

\bibitem{lin2010measuring}
Y.~Lin, K.~C. Desouza, and S.~Roy.
\newblock Measuring agility of networked organizational structures via network
  entropy and mutual information.
\newblock {\em Applied Mathematics and Computation}, 216(10):2824--2836, 2010.

\bibitem{sole2004information}
R.~V. Sol{\'e} and S.~Valverde.
\newblock Information theory of complex networks: on evolution and
  architectural constraints.
\newblock In {\em Complex Networks}, pages 189--207. Springer, 2004.

\bibitem{ji2008network}
L.~Ji, W.~Bing-Hong, W.~Wen-Xu, and Z.~Tao.
\newblock Network entropy based on topology configuration and its computation
  to random networks.
\newblock {\em Chinese Physics Letters}, 25(11):4177, 2008.

\bibitem{cover2012elements}
T.~M. Cover and J.~A. Thomas.
\newblock {\em Elements of information theory}.
\newblock John Wiley \& Sons, 2012.

\bibitem{Erdos1959random}
P.~Erd{\"o}s and A.~R{\'e}nyi.
\newblock On random graphs, i.
\newblock {\em Publicationes Mathematicae (Debrecen)}, 6:290--297, 1959.

\bibitem{knuth1992two}
D.~E. Knuth.
\newblock Two notes on notation.
\newblock {\em The American Mathematical Monthly}, 99(5):403--422, 1992.

\bibitem{wu15multinet}
H.~Wu.
\newblock Multinet.
\newblock \url{http://github.com/wuhaochen/multinet}.

\bibitem{dit}
R.~G. James, C.~J. Ellison, and J.~P. Crutchfield.
\newblock {dit}: Discrete information theory in {Python}.
\newblock \url{https://github.com/dit/dit}, 2018.

\bibitem{Goebel05}
B.~Goebel, Z.~Dawy, J.~Hagenauer, and J.~C. Mueller.
\newblock An approximation to the distribution of finite sample size mutual
  information estimates.
\newblock In {\em Communications, 2005. ICC 2005. 2005 IEEE International
  Conference on}, volume~2, pages 1102--1106. IEEE, 2005.

\bibitem{gerner2002conflict}
D.~J. Gerner, P.~A. Schrodt, O.~Yilmaz, and R.~Abu-Jabr.
\newblock Conflict and mediation event observations (cameo): A new event data
  framework for the analysis of foreign policy interactions.
\newblock {\em International Studies Association, New Orleans}, 2002.

\bibitem{IcewsData}
E.~Boschee, J.~Lautenschlager, S.~O'Brien, S.~Shellman, J.~Starz, and M.~Ward.
\newblock {ICEWS} coded event data, 2016.

\bibitem{AirData}
T-100 domestic segment.
\newblock \url{http://www.transtats.bts.gov/Fields.asp?Table_ID=259}.

\bibitem{ito2005assessing}
H.~Ito and D.~Lee.
\newblock Assessing the impact of the september 11 terrorist attacks on us
  airline demand.
\newblock {\em Journal of Economics and Business}, 57(1):75--95, 2005.

\bibitem{UNComtrade}
G.~Gaulier and S.~Zignago.
\newblock Baci: International trade database at the product-level. the
  1994-2007 version.
\newblock Working Papers 2010-23, CEPII, 2010.

\bibitem{leeds2002alliance}
B.~Leeds, J.~Ritter, S.~Mitchell, and A.~Long.
\newblock Alliance treaty obligations and provisions, 1815-1944.
\newblock {\em International Interactions}, 28(3):237--260, 2002.

\bibitem{williams2010nonnegative}
P.~L. Williams and R.~D. Beer.
\newblock Nonnegative decomposition of multivariate information.
\newblock {\em arXiv:1004.2515}, 2010.

\bibitem{stark2006biogrid}
C.~Stark, B.~Breitkreutz, T.~Reguly, L.~Boucher, A.~Breitkreutz, and M.~Tyers.
\newblock Biogrid: a general repository for interaction datasets.
\newblock {\em Nucleic Acids Research}, 34(suppl\_1):D535--D539, 2006.

\bibitem{duval1980reconsidering}
R.~D. Duval and W.~R. Thompson.
\newblock Reconsidering the aggregate relationship between size, economic
  development, and some types of foreign policy behavior.
\newblock {\em American Journal of Political Science}, pages 511--525, 1980.

\end{thebibliography}

\end{document}